\documentclass[manuscript,screen]{acmart}

\usepackage{color}
\usepackage{url}
\usepackage{hyperref}
\usepackage{algorithm,algcompatible}
\usepackage{algpseudocode}
\usepackage{amsmath}
\usepackage{float}
\usepackage{multirow}
\usepackage{booktabs}
\usepackage{xcolor,colortbl}

\usepackage{natbib} % to cite papers by authorname
\newcommand{\tcite}[1]{\citeauthor{#1}~\cite{#1}}

\newcommand{\runin}[1]{\noindent\textbf{#1.}}

\newcommand{\as}{adaptive system}

\AtBeginDocument{%
  }

\setcopyright{acmcopyright}
\copyrightyear{2022}
\acmYear{2022}
\acmDOI{XXXXXXX.XXXXXXX}

\acmBooktitle{ACM TAAS}
\begin{document}

\title{A User Study on Explainable Online Reinforcement Learning for Adaptive Systems}

\author{Andreas Metzger}
\authornote{Corresponding author.}
\email{andreas.metzger@paluno.uni-due.de}
\orcid{0000-0002-4808-8297}
\affiliation{%
  \institution{paluno (Ruhr Institute for Software Technology), University of Duisburg-Essen}
 \city{Essen}
  \country{Germany}
}

\author{Jan Laufer}
\email{jan.laufer@paluno.uni-due.de}
\orcid{0000-0002-3339-1760}
\affiliation{%
  \institution{paluno (Ruhr Institute for Software Technology), University of Duisburg-Essen}
 \city{Essen}  \country{Germany}
}

\author{Felix Feit}
\email{f.m.feit@gmail.com}
\orcid{0009-0006-2585-0062}
\affiliation{%
  \institution{paluno (Ruhr Institute for Software Technology), University of Duisburg-Essen}
 \city{Essen}  \country{Germany}
}

\author{Klaus Pohl}
\email{klaus.pohl@paluno.uni-due.de}
\orcid{0000-0003-2199-5257}
\affiliation{%
  \institution{paluno (Ruhr Institute for Software Technology), University of Duisburg-Essen}
   \city{Essen}\country{Germany}
}

\renewcommand{\shortauthors}{Metzger et al.}

\begin{abstract}
Online reinforcement learning (RL) is increasingly used for realizing adaptive systems in the presence of design time uncertainty. Online RL facilitates learning from actual operational data and thereby leverages feedback only available at runtime. However, Online RL requires the definition of an effective and correct reward function, which quantifies the feedback to the RL algorithm and thereby guides learning. With Deep RL gaining interest, the learned knowledge is no longer explicitly represented, but is represented as a neural network. For a human, it becomes practically impossible to relate the parametrization of the neural network to concrete RL decisions. Deep RL thus essentially appears as a black box, which severely limits the debugging of adaptive systems. We previously introduced the explainable RL technique XRL-DINE, which provides visual insights into why certain decisions were made at important time points. Here, we introduce an empirical user study involving 54 software engineers from academia and industry to assess (1) the performance of software engineers when performing different tasks using XRL-DINE and (2) the perceived usefulness and ease of use of XRL-DINE. 
\end{abstract}





\maketitle

\section{Introduction}
\label{sec:Introduction}

An \as{} (a.k.a. self-adaptive system) can modify its own structure and behavior at runtime based on its perception of the environment, of itself and of its requirements~\cite{Weyns2020}.
Examples of adaptive systems include elastic cloud systems~\cite{palm_online_2020}, intelligent IoT systems~\cite{ferry2021devops}, and proactive process management systems~\cite{metzger_triggering_2020}.

One key element of an \as\ is its \emph{adapt\-ation logic} that encodes when and how the system should adapt itself.
When developing the adaptation logic, developers face the challenge of \emph{design time uncertainty}~\cite{CalinescuMPW20,MetzgerEtAl2022,WeynsEtAl22}.
To define \emph{when} the system should adapt, they have to anticipate all potential environment states.
However, this is infeasible in most cases due to incomplete information at design time. 
As an example, the concrete services that may be dynamically bound during the execution of a service orchestration and thus their quality characteristics are typically not known at design time.
To define \emph{how} the system should adapt itself, developers need to know the precise effect an adaptation action has.
However, the precise effect may not be known at design time.
As an example, while developers may know in principle that enabling more features will negatively influence the performance, exactly determining the performance impact is more challenging.
A recent industrial survey identified optimal design and design complexity together with design time uncertainty to be the most frequently observed difficulties in designing adaptation in practice~\cite{WeynsEtAl22}.

\subsection{Online Reinforcement Learning for Adaptive Systems}
Online reinforcement learning (Online RL) is an emerging approach to realize \as{}s in the presence of design time uncertainty.
Online RL means that reinforcement learning~\cite{sutton_reinforcement_2018} is employed at runtime. 
A recent discussion of existing Online RL approaches for \as{}s is presented in ~\cite{MetzgerEtAl2022}.
Using Online RL, \as{}s can learn from actual operational data and thereby leverage information only available at runtime.
A recent survey indicates that since 2019 the use of learning dominates over the use of predetermined and static policies or rules~\cite{PorterFD20}.

Online RL aims at learning suitable adaptation actions via the \as{}'s interactions with its initially unknown environment~\cite{CAISE2020}. 
During system operation, the RL algorithm receives a numerical reward based on actual runtime monitoring data for executing an adaptation action.
The reward expresses how suitable that adaptation action was in the short term. 
The goal of Online RL is to maximize cumulative rewards. 

Earlier research on Online RL for \as{}s leveraged RL algorithms that represent learned knowledge as a so-called \textit{value function}~\cite{CAISE2020}.
The value function quantifies how much cumulative reward can be expected if a particular adaptation is chosen in a given environment state. 
Typically, this value function was represented as a table. 
However, such tabular approaches exhibit key limitations. 
First, they require a finite set of environment states and a finite set of adaptations and thus cannot be directly applied to continuous state spaces or continuous action spaces. 
Second, they do not generalize over neighboring states, which leads to slow learning in the presence of continuous environment states~\cite{CAISE2020}.

\textit{Deep reinforcement learning} (\emph{Deep RL}) addresses these disadvantages by representing the learned knowledge as a neural network. 
Since neural network inputs are not limited to elements of finite or discrete sets, and neural networks can generalize well over inputs, deep RL has shown remarkable success in different application areas. 
Recently, Deep RL is also being applied to \as{}s~\cite{moustafa2018deep,wang2019adaptive,CAISE2020}.
%One particular benefit of Deep RL for \as{}s is that it can capture concept drift in \as{}s without the need to explicitly introduce mechanisms to observe such drift~\cite{BPM2020,CAISE2020}

\subsection{Explainable Online Reinforcement Learning}

Understanding the decisions made by Deep RL systems is key to (1) increase the users' trust in these systems, and (2) help developers perform debugging~\cite{miller_explanation_2019,Dusparic_2022}.
Debugging of Deep RL is especially relevant for \as{}s, because Online RL does not completely eliminate manual development effort.
In particular, developers need to explicitly define a reward function, which 
quantifies the feedback to the RL algorithm.
Such a reward function may be derived from a utility function that balances the various, often conflicting, goal dimensions.
Getting such a utility function "right" -- in a sense that it accurately reflects the trade-off among different goal dimensions -- is a challenge~\cite{Dewey14}.
Additionally, recent findings show that modeling the reward function too closely on reality may slow down learning~\cite{BPM2020}.
As a consequence, defining the reward function introduces a potential source for human error.

A principal problem for debugging Deep RL is that the decision-making of the RL agent is not transparent, which implies the following main difficulties:
\begin{itemize}
\item 
Deriving the dynamic behavior of an RL agent from its reward function is hard, because reward functions may be rather complex; e.g., they may balance multiple competing goals~\cite{CAISE2020,ACSOS22} or include strategies such as artificial curiosity~\cite{PathakAED17}.
\item It is impossible to deduce the internal decision-making of an RL agent of a realistic complexity by only observing its interaction with the environment, i.e., by observing the evolution of states, actions and rewards.
\item The learned knowledge (and thus decision-making policy) of the RL agent is not represented explicitly.
Instead, it is ``hidden'' in the parametrization of the neural network. 
For a human, it is practically impossible to relate this parametrization to concrete RL decisions.
\end{itemize}

\begin{figure}[hbtp]
\centering
\includegraphics[width=.70\textwidth]{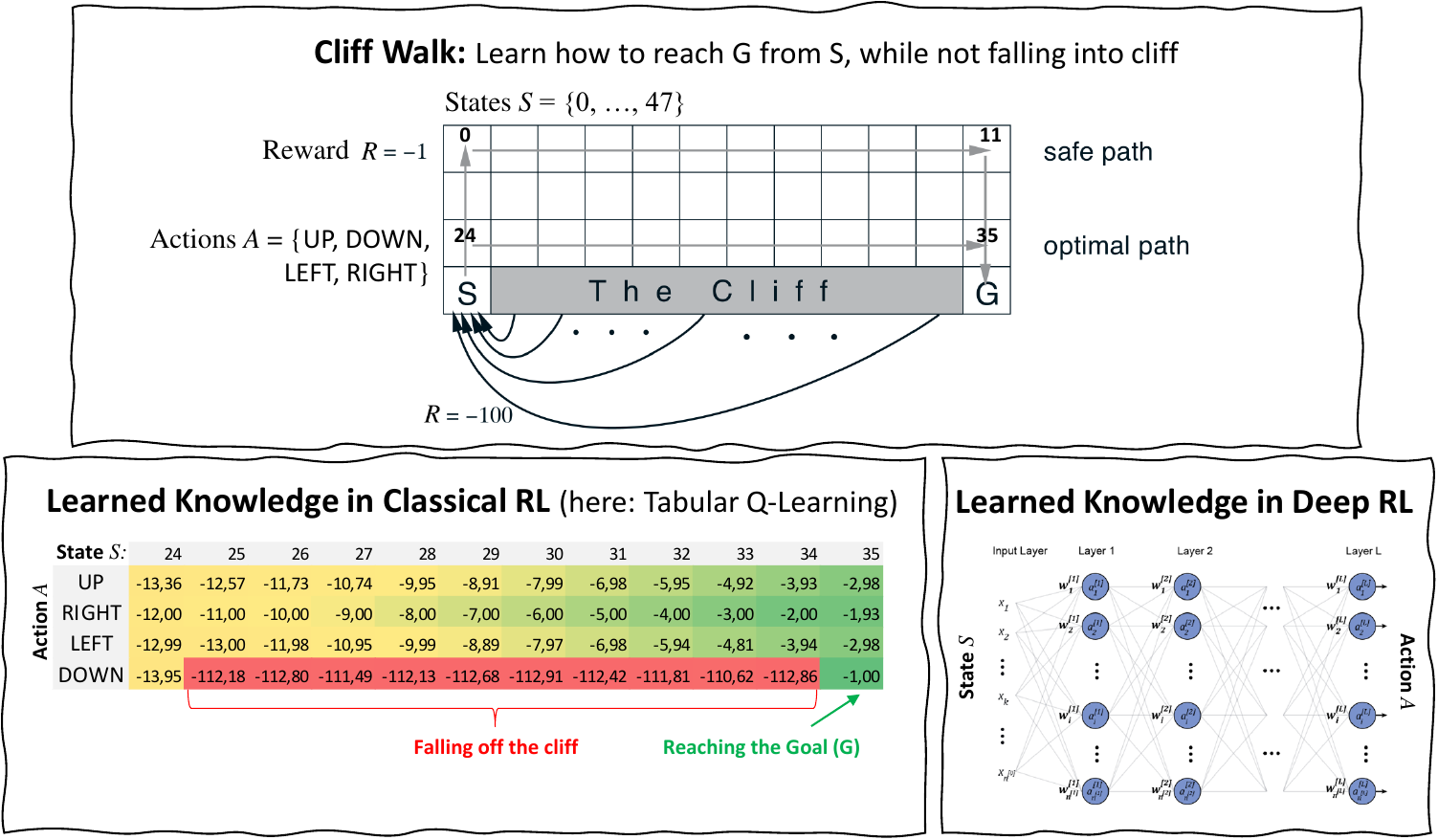}
\vspace{-.5em}
\caption{Illustration of how learned knowledge is represented for cliff walk example from~\cite{sutton_reinforcement_2018}.}
\vspace{-.5em}
\Description[<short description>]{<long description>}
\label{fig:problem}
\end{figure}

As a result, the decision-making of Deep RL essentially appears as a black box~\cite{puiutta_explainable_2020}.
Fig.~\ref{fig:problem} illustrates this problem using the classical RL toy example of cliff walk.
While it can clearly be seen that tabular Q-Learning has learned to avoid falling off the cliff and to reach the goal in an optimal way, this is not evident by looking at the weights of the neural network for Deep RL.

In our previous work~\cite{ACSOS22}, we introduced the \emph{XRL-DINE} technique that provides detailed explanations of RL decisions at relevant points in time.
XRL-DINE enhances and combines two existing explainable RL techniques from the machine learning literature: \textit{Reward Decomposition}~\cite{juozapaitis_explainable_2019} and \textit{Interestingness Elements}~\cite{sequeira_interestingness_2020}.
Reward Decomposition uses a suitable decomposition of the reward function into sub-functions to explain the short-term goal orientation of RL, thereby providing contrastive explanations.
Reward composition is especially helpful for the typical problem of adapting a system while taking into account multiple quality goals.
Each of these quality goals could then be expressed as a reward sub-function.
These explanations help to understand which goal an adaptation chosen by RL contributes to.
However, no indication for the explanation's relevance is provided, but instead it requires manually selecting relevant RL decisions to be explained. 
In particular when RL decisions are taken at runtime, which is the case for Online RL for \as{}s, monitoring all explanations to identify relevant ones introduces cognitive overhead for developers.
In contrast, Interestingness Elements collect and evaluate metrics at runtime to identify relevant moments of interaction between the system and its environment. 
Interestingness Elements thereby facilitate selecting relevant actions.
However, for an identified relevant moment of interaction, it does not provide explanations whether the system's decision making behaves as expected and due to the right reasons.

\subsection{Problem Statement and Contributions}
To assess the applicability and potential usefulness of XRL-DINE, we performed an initial evaluation in our previous work~\cite{ACSOS22}.
First, we prototypically implemented XRL-DINE using a state-of-the-art deep RL algorithm, serving as proof-of-concept.
Second, we demonstrated the potential usefulness of XRL-DINE by applying it to an adaptive web application exemplar~\cite{moreno_swim_2018}.
Third, we measured indicators for the potential reduction in cognitive load required to interpret explanations depending on how XRL-DINE was configured.

This initial evaluation indicated that XRL-DINE may be used by developers to gain insight and spot errors in the decision-making process of RL-based \as{}s.
However, our initial evaluation did not directly take into account the "human factor"~\cite{Doshi-VelezK17}, meaning that we did not involve actual developers in our evaluation process.

Our main new contribution is a systematic empirical user study that considers the "human factor" for evaluating XRL-DINE following the human-grounded evaluation approach proposed by \cite{Doshi-VelezK17}.
In this user study, we involved 54 software engineers from academia and industry.
In particular, we evaluate XRL-DINE considering three main research questions:
\begin{itemize}
\item \emph{What is the efficiency and effectiveness of software engineers when performing different tasks using XRL-DINE?}
\item \emph{What is the perceived usefulness and ease of use of XRL-DINE?}
\end{itemize}

In addition, we provide concise definitions and clearly differentiate the concepts "explanation", "explainability", and "interpretability", which are often used interchangeably in the literature.
This helps to more clearly position the contributions of XRL-DINE, and in particular serves as a conceptual foundation for our empirical user study.

Finally, we augment the original XRL-DINE technique by providing textual explanations in the form of counterfactuals to support explainability.

\subsection{Paper Organization}
Sect.~\ref{sec:Foundations} provides foundations as basis for introducing XRL-DINE in Sect.~\ref{sec:Approach}.
Sect.~\ref{sec:poc} describes the proof-of-concept implementation of XRL-DINE.
Sect.~\ref{sec:application} demonstrates the application of XRL-DINE to an adaptive systems exemplar.
Sect.~\ref{sec:user-study} describes the design, execution and results of our user study. Sect.~\ref{sec:discussion} discusses limitations.
Sect.~\ref{sec:RelatedWork} relates XRL-DINE to existing work.

\section{Foundations}
\label{sec:Foundations}

Below, we provide relevant foundations on online RL for adaptive systems, as well as explainable machine learning.

\subsection{Online RL for Adaptive Systems}

\subsubsection{Reinforcement Learning (RL)} 
\begin{figure}[hbtp]
\centering
\includegraphics[width=.8\textwidth]{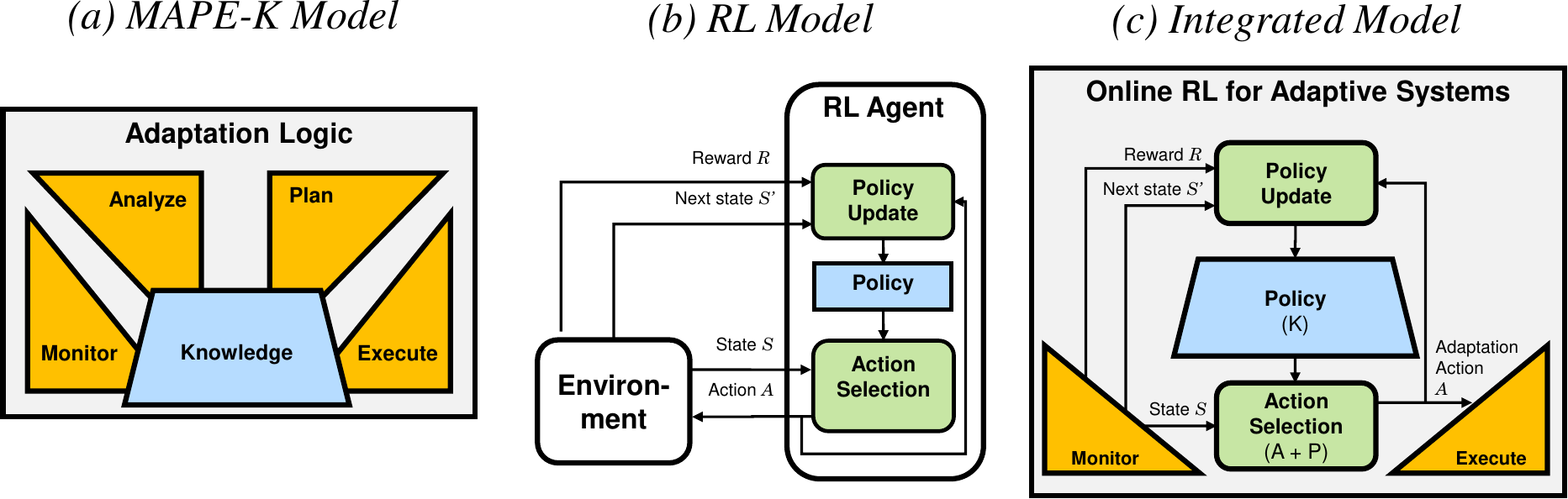}
\vspace{-.5em}
\caption{RL, MAPE-K, and their integration (adapted from~\cite{MetzgerEtAl2022}) }
\vspace{-.5em}
\Description[<short description>]{<long description>}
\label{fig:architecture}
\end{figure}

RL aims to learn an optimal action selection policy for a system -- called agent in RL.
To do so, an RL agent interacts with an initially unknown environment~\cite{sutton_reinforcement_2018}.
As  shown in Fig.~\ref{fig:architecture}\emph{(a)}, the 
agent finds itself in environment state $s$ at a given time step.
The agent then selects an action $a$ (from its set of potential adaptation actions) and executes it. 
As a result, the environment transitions to the next state $s'$ 
and the agent receives a reward $r$ for executing the action.
The reward $r$ together with the information about the next state $s'$ are used to update the action selection policy of the agent.
The goal of RL is to maximize the cumulative reward.

XRL-DINE is capable of providing insights into the decision-making of a particular kind of Deep RL algorithms, so called value-based Deep RL (also see the discussion in Sect.~\ref{sec:discussion}).
In value-based Deep RL, the action selection policy depends on a learned action-value function, also called $Q$ function, $Q(S,A)$.
The action-value function gives the expected cumulative reward when executing adaptation action $A$ in state $S$.
Value-based Deep RL uses a neural network to approximate $Q(S,A)$~\cite{sutton_reinforcement_2018}.

\subsubsection{Adaptive systems}
An adaptive system can conceptually be structured into two main elements~\cite{Weyns2020}: the \emph{system logic} and the \emph{adaptation logic}.
To understand how RL can be leveraged for realizing the adaptation logic, we use the well-established MAPE-K reference model for adaptive systems~\cite{Weyns2020}.
As depicted in Fig.~\ref{fig:architecture}\emph{(b)}, MAPE-K structures the adaptation logic into four main conceptual activities that rely on a common \emph{knowledge} base.
These activities \emph{monitor} the system and its environment, \emph{analyze} monitored data to determine adaptation needs, \emph{plan} adaptation, and \emph{execute} these adaptations at runtime.

\subsubsection{Online RL for adaptive systems} 
Online RL applies RL during system operation, where actions have an effect on the live system, resulting in reward signals based on actual monitoring data~\cite{CAISE2020,MetzgerEtAl2022}.
Fig.~\ref{fig:architecture}\emph{(c)} depicts how the elements of RL are integrated with  MAPE-K to facilitate learning effective adaptations at runtime.
%For a self-adaptive system, ``agent'' refers to the adaptation logic of the system and ``action'' refers to an adaptation action~\cite{MetzgerEtAl2022}.
In the integrated model, \emph{action selection} of RL takes the place of the \emph{analyze} and \emph{plan} activities of MAPE-K.
The learned \emph{policy} takes the place of the adaptive system's \emph{knowledge} base. 
At runtime, the policy is used by the adaptation logic to select  an adaptation action $a$ based on the current state $s$ determined by \emph{monitoring}.
Action selection determines whether there is a need for an adaptation (given the current state) and plans (i.e., selects) the respective adaptation action to \emph{execute}.
Using the policy, either a specific adaptation actions is selected and then \emph{executed}, or no adaptation is executed, which means the system is left in its current state.

\subsection{Explainable Machine Learning}
\label{subsec:explainableML}
Explainability is becoming more and more important with the increasing use of deep learning for decision making~\cite{miller_explanation_2019,GuidottiMRTGP19,SamekM19}.
While deep learning outperforms traditional ML models in terms of accuracy, one major drawback is that deep learning models essentially appear as a black box to its users and developers.
This means that users and developers are not able to interpret a black-box ML model's outcome, i.e., its decision, prediction, or action.
Using such black-box models for decision-making entails significant risks~\cite{Rudin19}.

Below, we provide definitions of the concepts "explanation", "explainability", and "interpretability" to serve as foundations for the remainder of this paper.
Note, that there is not yet a consistent understanding and use of these terms.
As an example, explainability and interpretability are two related but distinct concepts in AI, but often used interchangeably in the literature.

\runin{Explanation}
The term "explanation" in explainable machine learning may refer to two concepts~\cite{miller_explanation_2019}:
($i$) The \emph{process} of explaining an ML outcome, and ($ii$) the \emph{result} of this process.
Explanation as a process in turn involves (a) a cognitive process that determines the causes for an observed ML outcome, and (b) a knowledge transfer process from the explainer to an explainee (a.k.a. listener).

\runin{Explainability}
Explainabilty refers to the capability of an ML model to provide causes for its outcomes~\cite{WeldB19}.
Explainability thus supports the aforementioned cognitive process.
In other words, explainability characterizes the degree to which the cause for an ML model outcome can be understood.

\runin{Interpretability}
A stronger concept than explainability is interpretability.
Interpretability refers to the to the ease with which a human can understand the internal workings of an ML model~\cite{WeldB19}. 
%This means interpretability characterizes the degree to which human can compute the outcome of a ML model based on a given input.
Ideally, an interpretable ML model is one that is easy to understand and analyze, even for people who are not experts in the field of ML.
Typical examples for interpretable ML models are decision trees or linear regression.
Interpretability facilitates explainability, as understanding the internal workings of an ML model helps identify the causes for an ML model outcome.
%As interpretability requires exhibiting how a ML model works internally, interpretability typically is not attainable for black-box ML models.

Both interpretability and explainability strongly depend on the explainee's prior knowledge, experience and limitations~\cite{Ribeiro0G16}.
This means that while a random forest (i.e., an ensemble of $m$ decision trees) is interpretable in principle, in practice one faces the explainee's cognitive limitations, if, e.g., $m$ is large or the number of input features is high.

\section{The XRL-DINE Technique}
\label{sec:Approach}

The XRL-DINE technique provides insights into the decision making of Online RL, thereby making Online RL for adaptive systems explainable.
XRL-DINE combines two explainable ML techniques: Reward Decomposition~\cite{juozapaitis_explainable_2019} and Interestingness Elements~\cite{sequeira_interestingness_2020}.
XRL-DINE combines these techniques in such a way as to overcome their respective limitations.

The main information provided by XRL-DINE to its users are so called \emph{Decomposed Interestingness Elements} (DINEs).
DINEs facilitate explanations, but are not explanations per se (see our definitions from Sect.~\ref{sec:Foundations}).
Rather, DINEs allow users to understand the decision-making of an RL agent and thus help understand the causes behind an RL agent's actions.

XRL-DINE provides three different types of DINEs.
Each DINE focuses on a different aspect of the decision-making process of Online RL.
Below, we explain the different DINEs by summarizing the respective baseline approach,  explaining the extensions for XRL-DINE, and showing a visualization of the DINEs.
We conclude with a description of the XRL-DINE dashboard integrating the visualizations of the DINEs.

\subsection{``Reward Channel Dominance'' DINE} 
\runin{Baseline approach}
Originally proposed to improve learning performance, reward decomposition was exploited in~\cite{juozapaitis_explainable_2019} for the sake of explainability.
Reward Decomposition divides the reward function of the RL agent into several sub-functions, called \emph{reward channels}, which reflect a different aspect of the learning goal.
For each of the sub-functions a separate RL agent, which we call \emph{sub-agent}, is trained.
To select a concrete action,  an aggregated value-function is computed by accumulating the values for each of the actions proposed by the different reward channels. 
The resulting aggregated value-function is then used for action selection, while trade-offs in decision making made by the composed agent become observable via the reward channels. 

\runin{Extension for XRL-DINE}
We propose two types of ``Reward Channel Dominance'' DINEs: 
\emph{Absolute Reward Channel Dominance }gives the action-values $Q(S,A)$ of each sub-agent for a given state $S$. 
Since the action-values are not bound, the values can also be negative and may also vary widely between sub-agents. 

To better explain the actual contribution of each sub-agent to the aggregated decision, we convert the absolute values into relative ones, leading to the notion of \emph{Relative Reward Channel Dominance}.
To compute these relative values, for each sub-agent, the value of the worst action is subtracted from the action-values of all other actions. 
This results in (1) all values being positive, and (2) limiting the imbalance of contributions only to the \emph{decisive} portion. 

Since the conversion of absolute to relative dominance values may involve loss of information, both types of elements are provided to the developers as part of XRL-DINE.

\begin{figure}[hbtp]
\centering
\includegraphics[width=.90\textwidth]{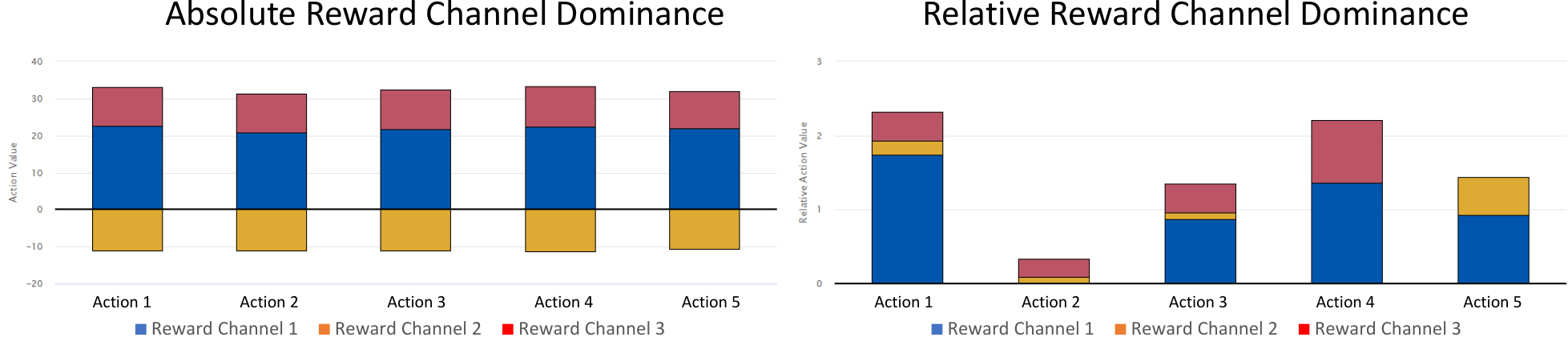}
\caption{Visualization of ``Reward Channel Dominance'' DINE for a given state $S$}
\Description[<short description>]{<long description>}
\label{fig:rcd}
\end{figure}

\runin{Visualization}
Fig.~\ref{fig:rcd} shows the visualization of these DINEs.
As can best be seen from relative reward channel dominance, Action 1 is chosen by the aggregated agent, as it has the highest relative rewards.
Here, reward channel 1 contributed most to the aggregate decision.
The same could also be seen from absolute reward channel dominance, but as the reward values are very close to each other, such a distinction is more difficult to make.

\subsection{``Uncertain Action'' DINE} 
\label{subsec:uncertainAction}
\runin{Baseline approach}
The ``Uncertain Action'' DINE is a modification of the Interestingness Element of \textit{certain} and \textit{uncertain} actions introduced in~\cite{sequeira_interestingness_2020}. 
The Interestingness Element of \textit{certain} and \textit{uncertain} actions can be used to reveal the RL agent’s confidence in its decisions.
Situations where the RL agent is uncertain of what to do may indicate parts of the state space in which the RL agent requires more training, or where the reward function may need to provide stronger rewards to improve decision-making of the RL agent,.

To determine whether an action is certain or uncertain,~\cite{sequeira_interestingness_2020} proposes measuring whether the RL agent in a given state chooses a wide range of different actions (and the RL agent is thus considered uncertain) or almost always chooses the same action (and the RL agent is thus considered certain). 
Given a state $S$, the execution certainty associated with
$S$ is measured by the concentration of the executions of the actions $A$. 

\runin{Extension for XRL-DINE}
This baseline approach is not directly applicable for value-based RL.
Value-based RL faces the exploration-exploitation dilemma~\cite{sutton_reinforcement_2018}. 
Actions should be selected that have shown to be effective (aka. exploitation). 
However, to discover such actions in the first place, actions that were not selected before should be selected (aka. exploration). 
One typical solution to the exploration-exploitation dilemma is the $\epsilon$-greedy mechanism. 
During learning, a random action is chosen with probability $\epsilon$ (exploration), while the action with the highest expected cumulative reward as determined by $Q(S,A)$ is chosen with probability $1-\epsilon$ (exploitation). 
To ensure convergence of the learning process, one challenge is to fine-tune the balance between exploitation and exploration ~\cite{CAISE2020}.
As an example, one may implement a mechanism that decreases $\epsilon$ over time, thereby reducing the amount of exploration  to facilitate convergence. 

In such a setting, the action chosen during exploitation (the so called \emph{greedy action}) is thus executed much more frequently than all other possible actions after convergence of the learning process.
This would cause the greedy action to be considered certain after convergence, even though other actions may only minimally differ from the greedy action in terms of expected cumulative reward as determined by $Q(S,A)$.

To overcome this weakness, we calculate the "Uncertain Action" DINE as follows.
We consider an action \textit{uncertain} if its action-value is very different \emph{relative} to the action-values of all alternative actions. 
This means we do not consider absolute action-values $Q(S,A)$, as they may be misleading if the agent is in a disproportionately good (or bad) part of the state-action space, where there is a naturally high (or low) expected cumulative reward $Q(S,A)$. 
Calculating the relative difference among action-values $Q(S,A)$ for all actions $A$ in a given state $S$ captures the internal relative weighting of the RL agent.

To determine whether an action-value is very different, we introduce a threshold parameter $\rho$.
This parameter prescribes how large the relative difference of action-values must be for a decision to be labeled as \textit{uncertain}. 
The lower this threshold is set, the more "Uncertain Action" DINEs will be generated. 
In contrast, for high thresholds, none or only a few, but possibly more relevant DINEs are generated.

To combine this with Reward Decomposition, the relative difference among action-values is calculated for each reward channel. 
Only if a difference is found for at least one of the reward channels and also the action of the aggregated RL agent does not correspond to the action that the sub-agent would choose, this identifies an "Uncertain Action". 

\begin{figure}[hbtp]
\centering
\includegraphics[width=.50\textwidth]{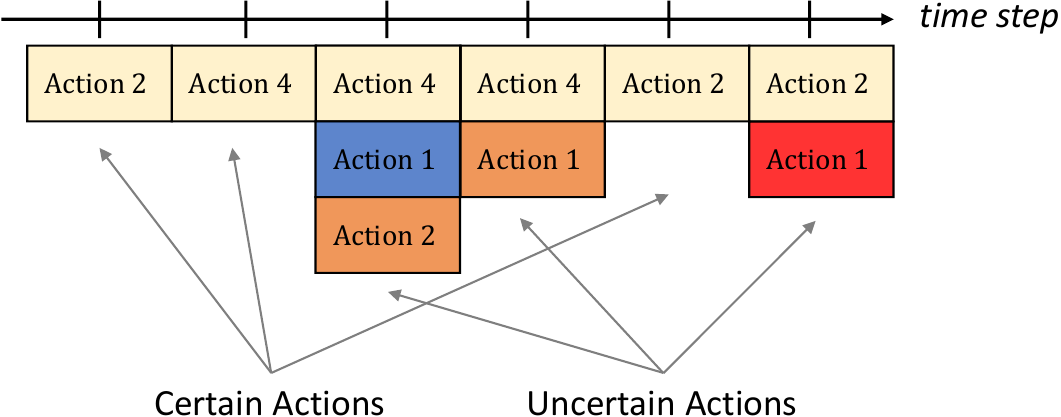}
\caption{Visualization of ``Uncertain Action'' DINE}
\Description[<short description>]{<long description>}
\label{fig:ia}
\end{figure}

\runin{Visualization}
Fig.~\ref{fig:ia} shows the visualization of these DINEs.
As can be seen, the first two RL decision at the beginning of the trace are certain, with the RL agent deciding for exactly one action.
This is followed by two uncertain actions, with one or even two alternative actions that may also have been chosen.

\vspace{-1em}

In addition to the graphical visualization, the "Uncertain Action" DINE also lends itself to generate contrastive explanations in which the action that would have been chosen by the sub-agent in isolation represents the contrastive action. 
XRL-DINE provides natural-language text for such an explanation following the below template: 

\begin{center}
\emph{To reach the goal <contrastive reward channel>, I should actually choose action <contrastive action>. 
However, it is currently more important to choose action <action chosen by aggregated RL agent> to achieve the goal <reward channel that dominated aggregated decision>. 
}\end{center}

%\noindent In this template, the first gap must be filled with the name of the reward channel corresponding to the subagent for which the action had high importance. 
%In addition, the second gap must be filled with the contrastive action preferred by this agent. 
%The third gap must then be filled with the actual action chosen by the aggregated agent, while the fourth gap must contain the name of the subagent that contributed most to the choice of the aggregated action.

\subsection{``Reward Channel Extremum'' DINE}
\runin{Baseline approach}
This DINE is based on the "Minima / Maxima Situations" of Interestingness Elements indicating local extrema of the state-value function $V(S)$, which gives the expected reward in state $S$. 
These extrema represent actions directly followed only by states with a higher state-value (local minimum) or lower state-value (local maximum)~\cite{sequeira_interestingness_2020}. 
Such local extrema  help identifying the RL agent's reasoning in potentially critical states~\cite{Dusparic_2022}.

\runin{Extension for XRL-DINE}
``Reward Channel Extremum'' DINEs can be linked to reward decomposition in two ways: 
First, a generic reward decomposition explanation can be displayed when the overall agent reaches a local maximum or minimum. 
Second, developers are informed when one of the sub-agents is in a local maximum or minimum. 

"Reward Channel Extremum" DINEs provide developers with an overview of how (1) the overall system and (2) each individual sub-agent evaluates a sequence of decisions and what actions are taken by the agent to leave a local reward minimum as quickly as possible or to maintain a high cumulative reward after a local reward maximum.

Even though the intuition behind ``Reward Channel Extremum'' DINEs is straightforward, deriving these DINEs requires predicting the next state $S'$ for each possible action $A \in \mathcal{A}$ to be able to determine maxima in an online fashion.
The value-based deep RL approaches we focus in this paper (see Sect.~\ref{sec:Foundations}) belong to the class of model-free RL algorithms.
This means that they do not use a model of the environment.
Such an environment model must therefore be approximated by XRL-DINE.
Such an approximation depends on the concretely chosen RL algorithm and is thus discussed in Sect.~\ref{sec:poc}. 

To generate the ``Reward Channel Extremum'' DINEs, at each time step for all possible actions $A \in \mathcal{A}$, and for the current state $S$ the next states are predicted using the approximated environment model. 
For these $|\mathcal{A}|$ predicted next states, all sub-agents then compute the state-value respectively. 
Since the sub-agents only approximate the action-value function, the state-value $V(S)$ is derived from the action-value function by choosing the action-value of the greedy action (i.e., the action with the highest $Q$ value):

\begin{equation}
V(S)=\max _{A \in \mathcal{A}} {Q}(S, A)
\end{equation}

The state value is calculated for the current state and all possible following states (derived from the environment model).
If the state-values of all predicted next states are worse than the state-values of the current state, a local reward maximum is reached. 
If all predicted next states are better, a local reward minimum is reached. 

Since local extrema may occur rather often, especially for a small number of discrete actions, we propose using a threshold $\phi$.
To determine a local minimum, the best action-value of the current state must be $\phi$ lower than the lowest action-value of the following state, and vice versa. 
The frequency and amount of information shown to the developers can thereby be controlled by $\phi$.

\begin{figure}[hbtp]
\centering
\includegraphics[width=.70\textwidth]{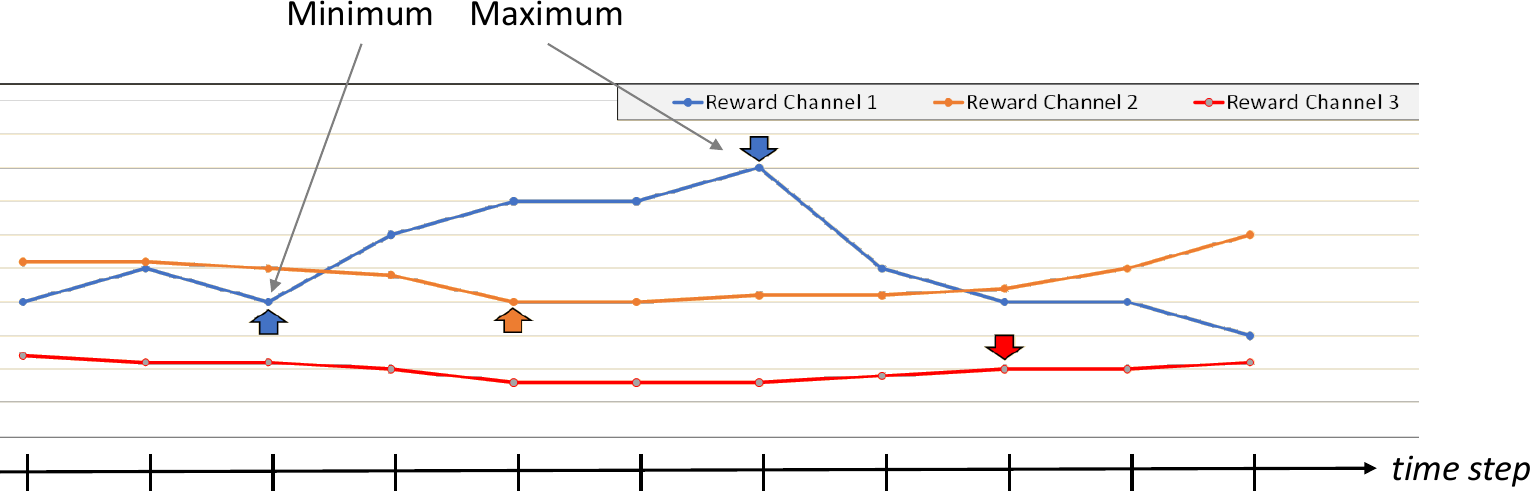}
\caption{Visualization of ``Reward Channel Extremum'' DINE}
\Description[<short description>]{<long description>}
\label{fig:re}
\end{figure}

\runin{Visualization} Fig.~\ref{fig:re} shows the visualization of these DINEs.
As can be seen, reward channel 1 rather quickly reaches a local maximum when compared to the other reward channels.
Yet, also rather quickly the reward in channel 1 drops -- even below the observed previous local minimum -- suggesting an important change in the RL agent's environment, which has not yet been sufficiently learned by reward channel 1.

\subsection{XRL-DINE Dashboard}

The DINEs are visualized and contextualized in the XRL-DINE dashboard. 
The dashboard allows navigating the decision trajectory and thereby also facilitates investigating explanations of past interactions. 
The purpose of this type of visualization is to preserve the respective advantages of the two combined explanation techniques:
Being able to gain an understanding of the RL agent's higher-level behavior, while also being able to investigate specific actions of interest.

\begin{figure*}[ht]
    \centering
%    	$
%		\begin{array}{cc}
    \includegraphics[width=.8\textwidth]{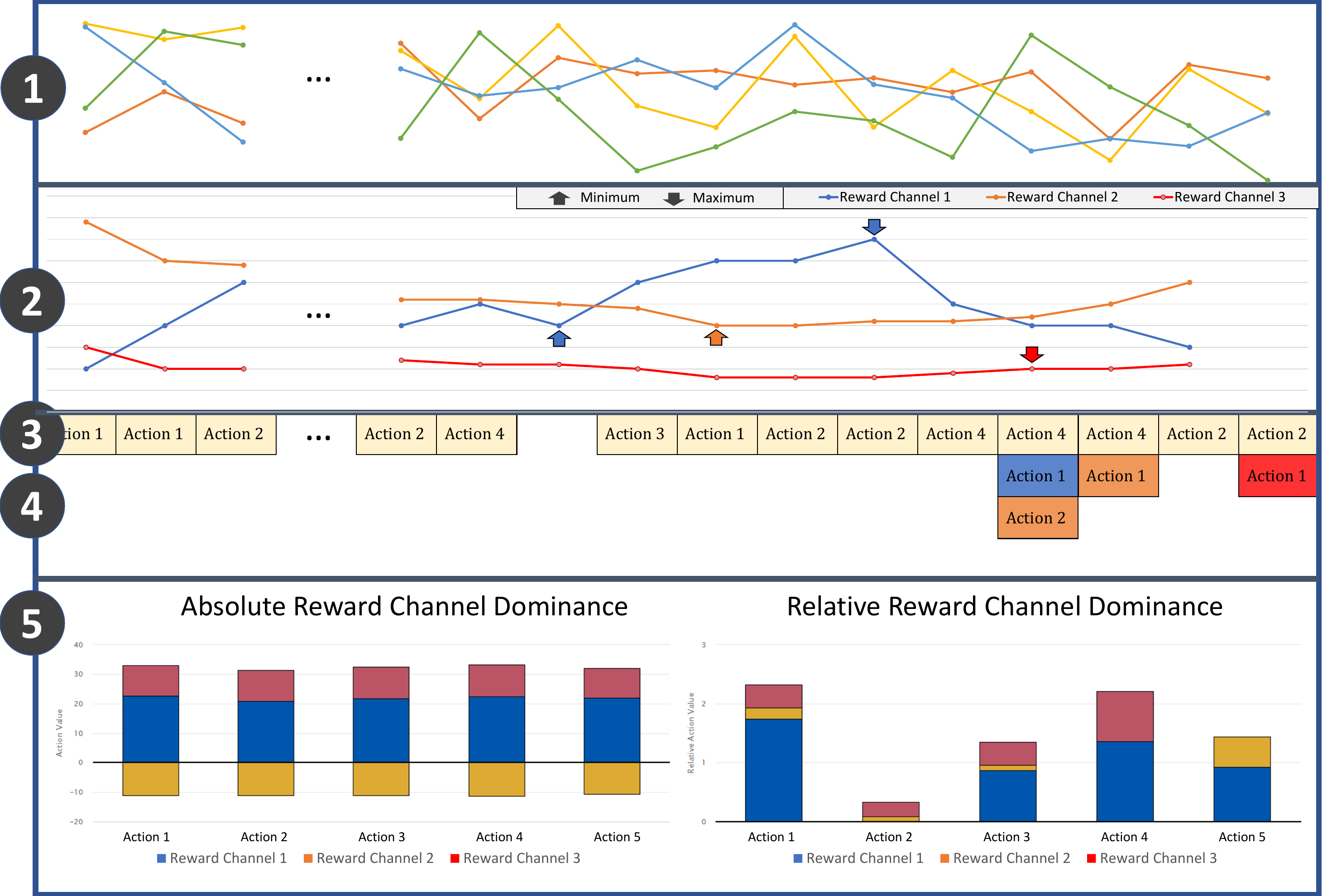}
%\end{array}$
    \caption{XRL-DINE dashboard}
\Description[<short description>]{<long description>}
    \label{fig:dashboard}
\end{figure*}

The dashboard is shown in Fig.~\ref{fig:dashboard}.
It  follows an interaction concept that allows visual data exploration~\cite{keim_visual_2001} and is centered around time progression as main axis. 
The different visual elements are linked, so that hovering over one element highlights the information of other elements for the same time step. 
By selecting a specific time step, the reward dominance values calculated for that time step and are displayed as barcharts. 
We took into account previous experience on user-centered design~\cite{endsley2003designing}.
To reduce cognitive overload, by (1) computing DINEs, and by (2) only showing relevant information when needed.
To avoid the requisite memory trap (i.e., relying on limited short-term memory), we show the complete historic information in a compact representation.

In detail, the XRL-DINE dashboard shows:
\begin{enumerate}
 \item \textbf{State progression}, which is represented using a line graph. 
Each line in the dashboard represents a Z-score standardized state variable. 
Standardization is necessary to display differently scaled variables. 
 \item \textbf{Received rewards progression} for each reward channel together with the ``Reward Channel Extremum'' DINEs.
 \item \textbf{Trajectory of selected actions}, where adaptation actions chosen by the composed RL agent are shown.
 \item \textbf{Uncertain Actions}, for which the background color corresponds to the color of the particular reward channel for which the contrastive action is considered important. 
If multiple reward channels suggest an action different from the selected action of the aggregated RL agent, they are stacked on top of each other.
 \item \textbf{Reward Channel Dominance}, which is displayed in a stacked column chart. 
Each column represents a possible adaptation action. 
The reward channels are shown in the same colors as for the other DINEs. 
 \end{enumerate}

\section{Proof of Concept}
\label{sec:poc}

\begin{figure*}[ht]
    \centering
    \includegraphics[width=1\columnwidth]{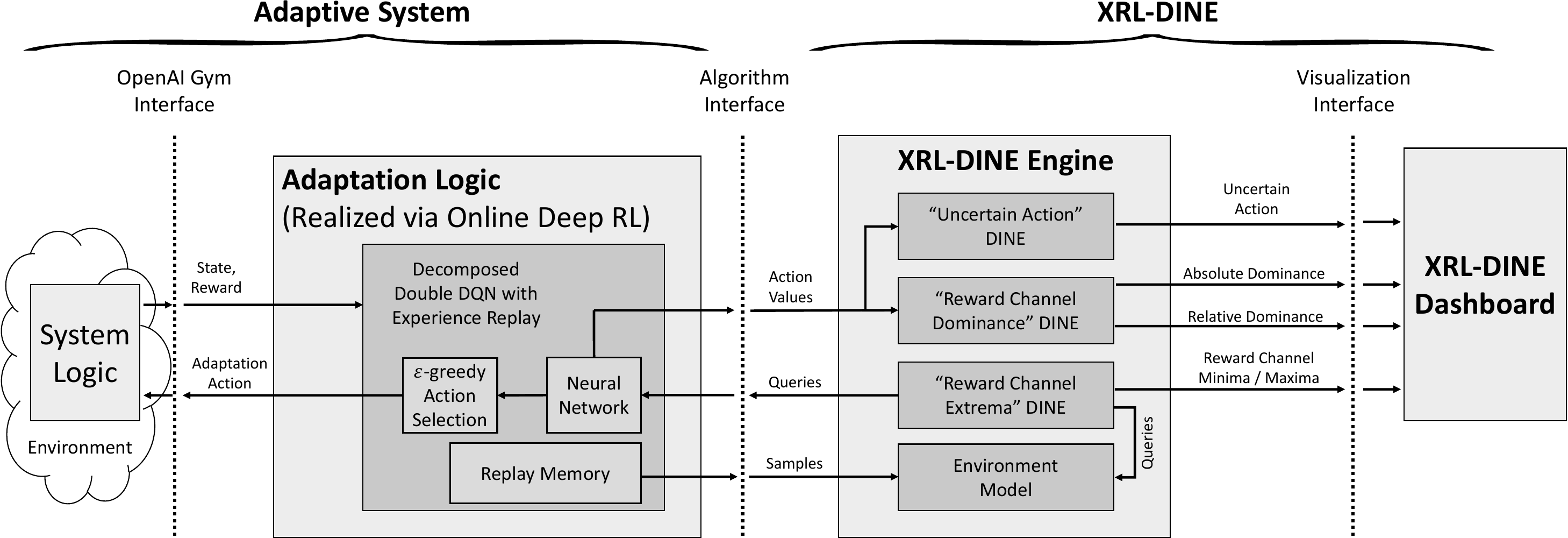}
    \caption{Architecture of Prototypical Implementation of XRL-DINE}
    \Description[<short description>]{<long description>}
    \label{fig:interfaces}
\end{figure*}

Following established practices in adaptive systems research~\cite{PorterFD20}, we perform real-world experiments to validate XRL-DINE.
This means we implement a system\footnote{Code is available via \url{https://git.uni-due.de/rl4sas/xrl-dine}} and then subject it to approximately realistic conditions.
Fig.~\ref{fig:interfaces} shows the main components of the XRL-DINE implementation (the XRL-DINE engine and dashboard) and how they connect with the implementation of the adaptive system using Deep RL.

We prototypically implemented XRL-DINE for \textit{Double Deep Q-Networks with Experience Replay} as a state-of-the-art variant of value-based deep RL algorithms~\cite{hasselt_deep_2016}. 
One particular benefit of this RL variant is that it addresses the problem of sparse rewards, i.e., that the RL agent only sporadically may receive rewards.
To this end, data observed during agent-environment interactions are not used directly for training, but are first written to a replay memory. 
At each time step, a predefined number of samples (called \textit{batch}) is randomly taken from the replay memory and the current bootstrapping target is calculated for each sample.
The bootstrapping target is an approximation of the expected cumulative reward of an action. 
It is formed as the weighted sum of the immediate reward received for the chosen action and the action-value of the best possible action in the successor state.
For the calculation of the bootstrapping target the other sub-agents are considered as well. 
We thus extend the Deep RL algorithm to the decomposed case with the update rule proposed in~\cite{juozapaitis_explainable_2019}. 
%Using the target and the predicted action-values we form the loss and then perform backpropagation to adjust the neural network weights to (gradually) minimize the loss.

As shown Fig.~\ref{fig:interfaces}, the Deep RL algorithm that realizes the adaptation logic is linked to the XRL-DINE Engine as follows.
First, the action values are transmitted to the XRL-DINE engine at each time step via a callback. 

Second, as explained in Sec.~\ref{sec:Approach}, an approximated environment model is needed for computing the ``Reward Channel Extremum'' DINEs.
In our implementation, we create such a model using supervised learning by exploiting the contents of the replay memory as a labeled dataset.
The replay memory contains past transitions in the form $(s, a, r, s')$ (i.e., State-Action-Reward-Next State~\cite{sutton_reinforcement_2018}). 
Using $s$ and $a$ as input and $s'$ as output, we train a generic feed-forward neural network as a universal nonlinear function approximator. 
During the initialization of the XRL-DINE engine, links to the replay memory (for training the environment model) and the neural networks of the individual RL sub-agents (for evaluating successor states) are passed by reference.
Of course, training an accurate environment model requires sufficient amount of training data.
Such training data is not available at the start of the RL learning process and it takes some time until sufficient data is collected.
Yet, this is not a critical concern for the applicability of XRL-DINE.
We are interested in explaining the RL agent's behavior after initial convergence of the learning process, which typically takes many iterations (e.g., in the order of several 10,000 iterations, like in our exemplar from Sect.~\ref{sec:application}) and thus a significant amount of data has been collected when needed.

This implementation of the adaptation logic is connected to the adaptive system and its environment using the OpenAI Gym interface, a widely used interface between RL agents and their environments\footnote{For details see \url{https://www.gymlibrary.ml/content/api/}}.
Classes that implement this interface offer a ``step'' method, to which the action to be executed is passed. 
This action is then executed and the next state and the reward received are returned. 
Since Reward Decomposition requires a vector of reward values instead of the default scalar reward value provided by the OpenAI Gym interface, we have slightly modified the interface.

\section{Application to Adaptive System Exemplar}
\label{sec:application}

We apply XRL-DINE to a concrete adaptive system exemplar\footnote{Supplementary information available via \url{https://git.uni-due.de/rl4sas/xrl-dine}} .
This serves as basis to demonstrate the use of XRL-DINE (Sect.~\ref{sec:apply}), measure indicators for the cognitive load of using XRL-DINE (Sect.~\ref{sec:quantitative}), and perform our user study (Sect.~\ref{sec:user-study}).

\subsection{SWIM Exemplar}
\label{sec:swim}
We chose SWIM as an exemplar, which is one of the adaptive system exemplars provided by the SEAMS community~\cite{moreno_swim_2018}. 
SWIM simulates an adaptive multi-tier web application.
It closely replicates the real-time behavior of an actual web application, while allowing to speed up the simulation to cover longer periods of real time.
SWIM instantiates the auto-scaling problem in which the objective is to provision resources to satisfy conflicting business goals. 
Specifically, adaptation logic must be implemented that maximizes a given utility function, despite varying system load.

SWIM has different monitoring metrics to determine the state of the system.
These metrics include the request arrival rate (i.e., ``workload'') as well as the average throughput and response time.
As all three environment variables are continuous, SWIM's state space is continuous.
Therefore, tabular RL solutions cannot directly be applied to the exemplar.

SWIM can be adapted as follows:
(1)~additional web servers can be added / removed, resulting in the load being distributed across more / fewer servers; (2)~the proportion of requests for which optional, computationally intensive content is generated (e.g., via recommendation engines) can be modified by setting a so called dimmer value. 
While adaptations of type (1) have an impact on costs, adaptations of type (2) have an  impact on revenue.

To apply XRL-DINE, we adapt a reward function from the literature~\cite{MorenoPACS17} and split it into three reward channels to form the following decomposed reward function:

\begin{equation}
R_{\mathrm{total }}=
a \cdot R_{\mathrm{user\_satisf.}} + 
b \cdot R_{\mathrm{revenue }} + 
c \cdot R_{\mathrm{costs }}
\end{equation}

The weights were selected experimentally according to two criteria. 
First, no reward channel should dominate the other two reward channels to such an extent that the decisions of the other sub-agents have no influence on the choice of actions. 
This criterion reflects the basic assumption of DINEs that multiple sub-agents should find the best possible trade-off. 
The second, subordinate criterion is to conform as closely as possible to the original utility function. 
Specifically, the parameters were chosen such that \textit{User Satisfaction} has the highest influence ($a = 4$), \textit{Revenue} has the second highest influence ($b = 2$), and \textit{Costs} has the lowest influence ($c = 1$). 

The three reward sub-functions are defined as follows:

\begin{equation}
R_{\text {user\_satisf.}}= \begin{cases}0.5 & : x \leq 0.02 \\ -0.5-\frac{x-1}{20} & : x \geq 1 \\ 0.5- \frac{x-0.02}{0.98} & : \text { otherwise }\end{cases}
\end{equation}

%\begin{align*}
%R_\textrm{user\_satisf.} = 0.5 &\textrm{ if } x \leq 0.02; \\
%-0.5-(x-1)/20 &\textrm{ if }  x \geq 1; \\
%0.5- (x-0.02)/0.98 &\textrm{ otherwise} 
%\end{align*}
In $R_\textrm{user\_satisf.}$,  the perceived user satisfaction depends on the average latency $x$.
This utility function is provided as part of the SWIM exemplar.

\begin{equation}
R_\textrm{revenue}= \tau \cdot a \cdot\left(d \cdot R_{O}+(1-d) \cdot R_{M}\right)
\end{equation}

In $R_\textrm{revenue}$, $\tau$ is the length of the time interval between two consecutive time steps, $a$ is the average arrival rate of requests and $d$ is the current dimmer value. 
$R_{M}$ is the reward obtained when processing a request without optional content. 
$R_{O}$ is the reward obtained when processing a request with optional content. 
The term $d \cdot R_{O}+(1-d) \cdot R_{M}$ thus represents the average reward controlled by the dimmer value for each request.

\begin{equation}
R_\textrm{costs} = -(\tau \cdot c \cdot s)
\end{equation}

In $R_\textrm{costs}$, $s$ indicates the number of servers currently in use. 
The parameter $c$ models the cost of using one server.
This means, $R_\textrm{costs}$ is higher the fewer servers are used.

Finally, 
%to prevent illegal actions, we explicitly punish them by adding a fixed penalty to each reward channel. 
all actions that cause a change in the system state (i.e., all actions except ``No Adaptation'') are penalized by a reward of $-0.1$ to account for increased computational overhead and to incentivize the aggregated RL agent to learn a  less jerky policy.

%We apply the \textit{Decomposed Double Deep Q-Learning Agent} explained above to the SWIM environment. 

\subsection{Demonstrating the Use of XRL-DINE}
\label{sec:apply}
To demonstrate the applicability of XRL-DINE and to illustrate the insights that DINEs provide, we selected a scenario from the SWIM exemplar.
Below, we explain how XRL-DINE may be used to gain insights into the decision making of the RL agent for this scenario.
Our chosen scenario\footnote{The interactive XRL-DINE dashboard for this scenario is available via \url{https://git.uni-due.de/rl4sas/xrl-dine}} begins at time steps after the initial convergence of the RL process, which happens at timestep $T + 1 =$ 22,575.
Using XRL-DINE to gain insights into the decision making of the RL agent after convergence allows software engineers  to determine whether the adaptive system behaves as expected and it may help uncover potential problems in how the reward function was defined.

\begin{figure*}[ht]
    \centering
    \includegraphics[width=.85\textwidth]{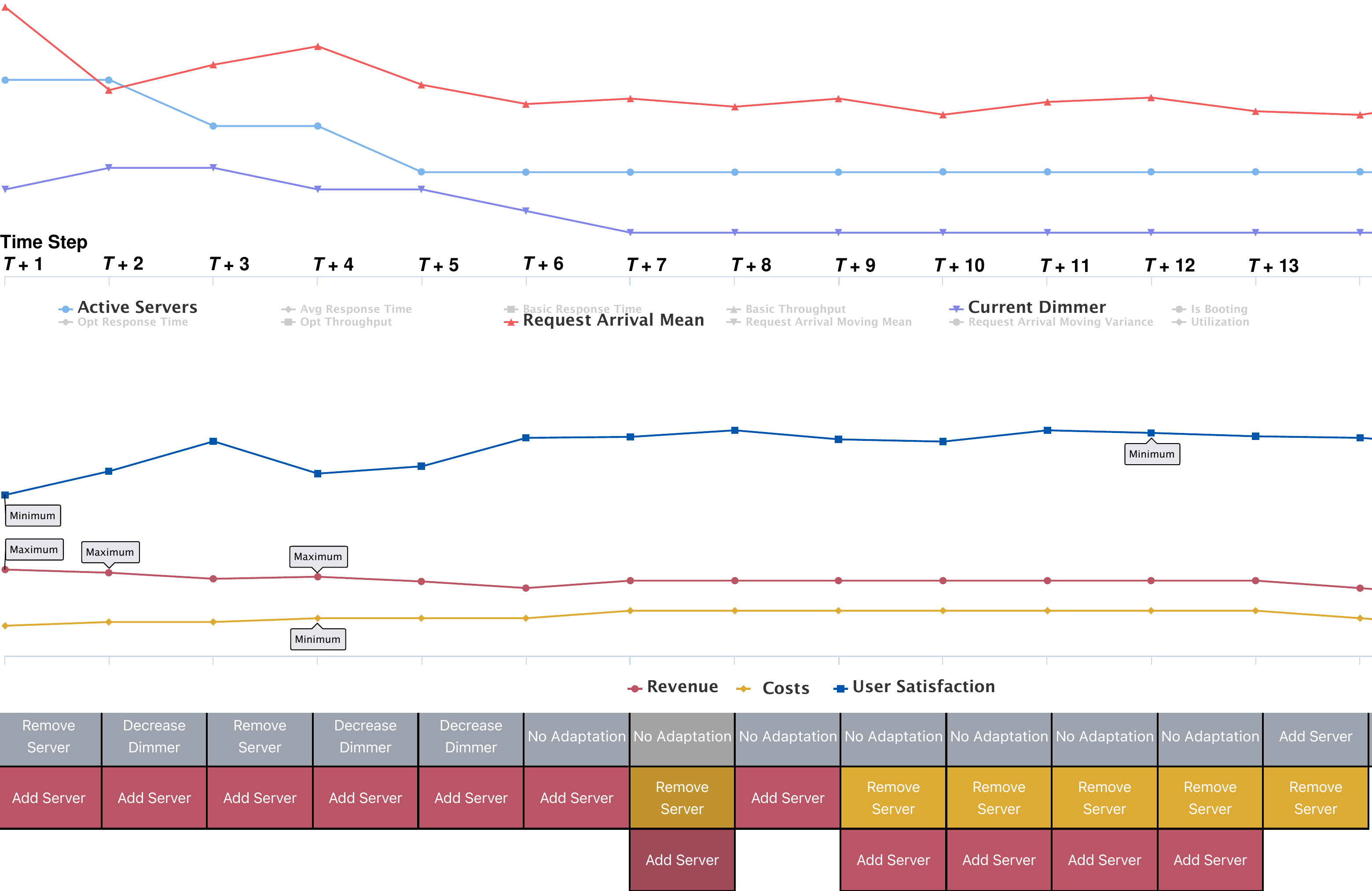} 
    \caption{XRL-DINE dashboard for scenario shown starting from time step $T + 1 = $ 22,575}
\Description[<short description>]{<long description>}
    \label{fig:scenario1}
\end{figure*}

Fig.~\ref{fig:scenario1} shows the contents of the main part of the XRL-DINE dashboard for this scenario.
The figure shows how the RL agent responds to an increase in user requests. 
This increase is reflected in the red curve in the XRL-DINE dashboard. 
This curve represents the normalized duration between incoming requests. 
The lower this value, the higher the request rate (i.e., the request rate is the inverse of the normalized duration between incoming requests). 
At time step $T + 2$, there is a short peak in the request rate, which then decreases again. 
Between time steps $T + 4$ and $T + 6$, the request rate increases again and then more or less remains at this higher level for the remainder of the example. 
The number of active servers is ten at time step $T + 1$, but is lowered by the RL agent down to eight servers by time step $T + 5$. 
The dimmer value is at 0.1 at time step $T + 1 $ and is reduced by the RL agent to 0.0 by time step $T + 7$.

The curve of the \textit{User Satisfaction} reward channel depicts an increase between time steps $T + 1$ and $T + 3$, which decreases again due to the shutdown of two servers after time step $T + 3$. 
By lowering the dimmer value in time steps $T + 4$ and $T + 6 $ and due to a more or less stable request rate, the reward for the \textit{User Satisfaction} reward channel increases and stabilizes at a high level from time step $T + 6$ onward. 
Lowering the dimmer value also causes the reward for the \textit{Revenue} channel to decrease and stabilize at a lower level. 
In contrast, the reward for \textit{Running Costs} increases, since two less servers need to be operated. 
By choosing \emph{No Adaptation} between time steps $T + 7$ and $T + 13$, all reward channels receive a relative boost, as the $-0.1$ penalty for choosing any action except \emph{No Adaptation} is no longer received. 
In summary, the aggregated RL agent's strategy is to respond to an increase in the request rate by lowering the dimmer, thus trading a gain in \textit{User Satisfaction} and \textit{Running Costs} for a loss of \textit{Revenue}.

Between time steps $T + 2$ and $T + 13$, the \emph{Revenue} sub-agent decides to activate more servers. 
Yet, this decision is never taken by the aggregated RL agent until time step $T + 13$. 
The \textit{Running Costs} sub-agent, in contrast, regularly decides to turn off more servers in the second half of the example, when the request rate drops again. 
This is in direct contradiction to the decision of the \textit{Revenue} sub-agent. 

\begin{figure*}[ht]
    \centering
    	\includegraphics[width=.6\textwidth]{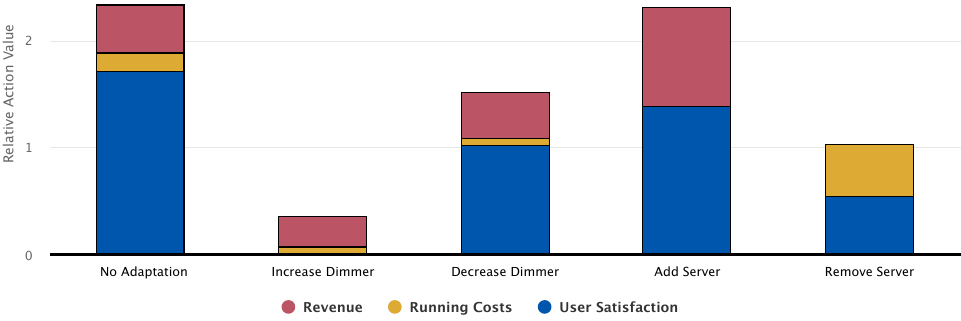} 		
    \caption{"Reward Channel Dominance" DINE for time step $T + 13$}
\Description[<short description>]{<long description>}
    \label{fig:scenario1rcd}
\end{figure*}

To better understand the aggregated agent's internal decision making when the aggregated decision changes to "Add Server", we look at the ``Reward Channel Dominance'' DINEs for time step $T + 13$, shown in Fig.~\ref{fig:scenario1rcd}.

The action taken at time step $T + 13$ is the last action before actually adding another server, thus satisfying the repeated request of the sub-agent for \textit{Revenue}. 
In addition to the \textit{Revenue} sub-agent, the \textit{Running Costs} sub-agent also proposes an alternate action at this time step.
The ``Relative Reward Channel Dominance'' DINE shows that the sub-agent for \textit{User Satisfaction} has the greatest influence on the selected action \textit{No Adaptation}. 
The other two sub-agents each show visibly less reward channel dominance.  
This imbalance suggests a strongly biased action selection. 
It can also be seen that the chosen action is closely followed by the second-best action \emph{Add Server}. 
The alternative action \emph{Remove Server} proposed by the \textit{Running Costs} sub-agent is significantly worse, with less than half as much cumulative Relative Reward Channel Dominance, compared to the \emph{No Adaptation} action.

%Regarding the "Reward Channel Extremum" DINE, we can observe: 
%At the beginning and at the end, the trace for the \textit{User Satisfaction} reward channel shows minima, which means that the sub-agent expects an increase in action-values in all possible subsequent states. 
%In contrast, the sub-agent for \textit{Revenue} expects a decrease in action-values in three of the first four steps. 
%The agent for \textit{Running Costs} determines a local minimum of action-values at the beginning and at the end of the sequence.

In the example, the aggregated agent decides to sacrifice \textit{Revenue} for higher \textit{User Satisfaction} and lower \textit{Running Costs}. Meanwhile, the sub-agent for \textit{Revenue} suggests alternative actions.
This is something one may expect when having some knowledge of the domain. 
These suggestions all relate to the \emph{Add Server} action. Adding more servers results in a lower average server utilization. 
This lower utilization would allow the dimmer to be raised again without compromising the dominant \textit{User Satisfaction} reward. 
Thus, from a domain perspective, these alternative actions make sense. 
The proposed alternative actions of the \textit{Running Costs} sub-agent also make sense from domain point of view, since removing servers intuitively leads to lower server costs.

Summarizing, DINEs in this example facilitate explicitly representing the RL agent's internal decision making trade-offs. 
From a domain point of view, one can think of three reasonable strategies for responding to the unanticipated increase in request rate shown in the example: (1) more servers can be added to ensure a high level of \textit{Revenue} and \textit{User Satisfaction} but at the expense of \textit{Running Costs}, (2) the dimmer value can be lowered to ensure a high level of \textit{User Satisfaction} and \textit{Running Costs} at the expense of \textit{Revenue}, or (3) no action at all can be taken at the expense of \textit{User Satisfaction}.

In the chosen example, the agent follows the second strategy. 
Software engineers can evaluate this strategy and, if they are dissatisfied with it (and may consider it as buggy), they may change the reward function such as to prefer the alternative strategy suggested by the Revenue sub-agent.  

To further support such debugging of the RL agent, ``Reward Channel Dominance'' DINEs can be leveraged to provide additional insights. 
For example, the developer can identify from the dashboard that \textit{User Satisfaction} alone is not sufficient for the agent to favor \emph{No Adaptation} over \emph{Add Server}. 
Instead, the relatively small influence of \textit{Running Costs} ultimately makes the difference. 
By investigating the ``Reward Channel Dominance'' DINEs of the selected actions, developers may determine that, from an agent's perspective, the first strategy listed above is rated second best by the agent.
This insight may then be used to adjust the structure or weighting of the reward function's components.

\subsection{Indicators for Cognitive Load of XRL-DINE}
\label{sec:quantitative}
\begin{figure*}[ht]
    \centering
    \includegraphics[width=.4\textwidth]{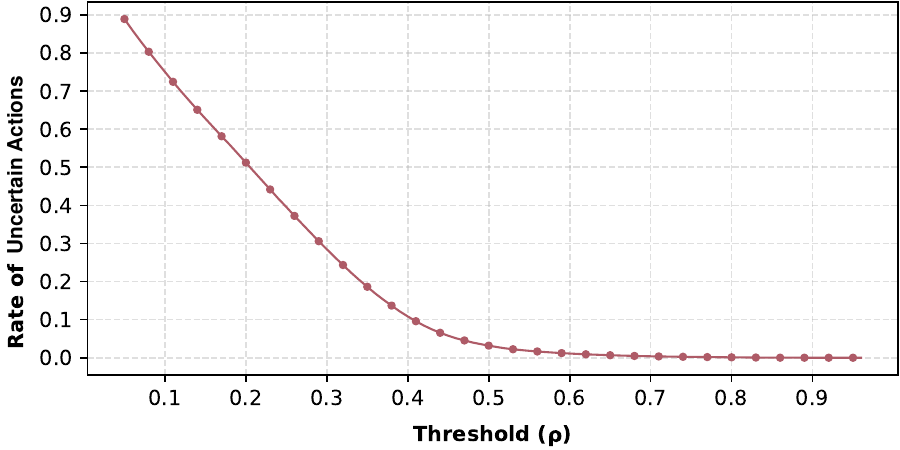}     
    
	$	\begin{array}{cc}
	\\
\includegraphics[width=.4\textwidth]{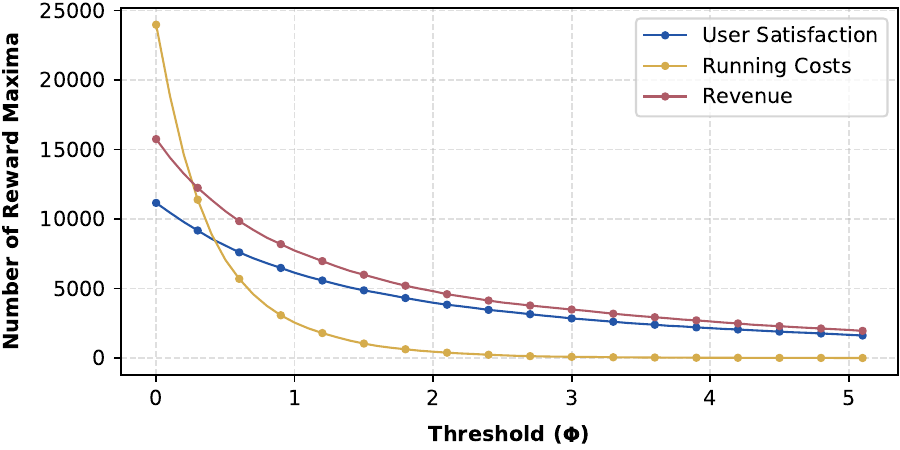} &		
	\includegraphics[width=.4\textwidth]{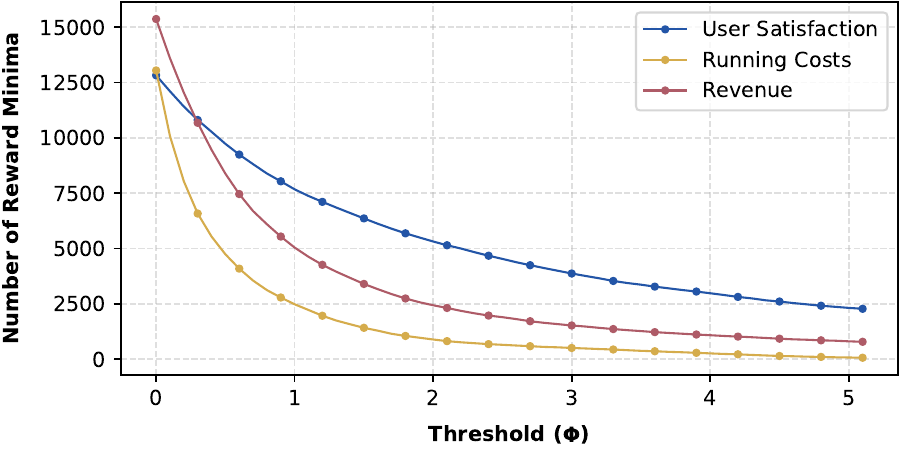}
\end{array}$
    \caption{Influence of threshold on number of DINEs shown in XRL-DINE dashboard}
\Description[<short description>]{<long description>}
    \label{fig:thresholds}
\vspace{-2em}
\end{figure*}

To complement the above results, we measure the number of DINEs shown to developers, thereby serving as indicator for the cognitive load.
We perform measurements for the different workload traces, covering a total of 62,000 timesteps.
In particular, we measure  how the number of DINEs depends on choosing the two thresholds: $\rho$ for ``Uncertain Action'' DINEs and $\phi$ for ``Reward Channel Extremum'' DINEs.
To facilitate comparability of results, we use data from a single run of the RL agent, but filter accordingly based on the different thresholds.
Results are shown in Fig.~\ref{fig:thresholds}.

As can be seen from the charts, the thresholds allow tuning the rate of DINEs in a wide range; e.g., from close to zero up to 100\% in case of the $\rho$ threshold.
As the thresholds can be changed at runtime, developers may dynamically tune the rate of DINEs (1) based on their needs; e.g., coarse-grain observation vs. in-depth debugging, and (2) depending on the frequency of environment interactions.

\section{Empirical User Study}
\label{sec:user-study}
In Sect.~\ref{sec:application}, we have illustrated how XRL-DINE may help to understand the decision making of the online RL agent.
This has highlighted the principal capabilities of XRL-DINE.
In this section, we provide empirical evidence in how far XRL-DINE indeed helps software engineers to perform tasks related to understanding the decision making of the online RL agent, as well as how useful software engineers perceive XRL-DINE to be.

\subsection{Research Questions}
\label{sec:rq}

We evaluate XRL-DINE considering the following research questions:
\begin{itemize}
\item \emph{RQ1: What is the performance of software engineers when performing different tasks using XRL-DINE?} 
The assessment of the quality of explanations is an active field of research~\cite{SamekM19}.
Different techniques for assessing the quality of explanations exist, which include asking for a summary of the given explanation in someone's own words, observing humans' behavior when performing distinct tasks, or measuring the human task performance when using explanations. 
In our user study we were interested in measuring task performance by asking participants to solve dedicated tasks using XRL-DINE. 
We use task performance in terms of effectiveness and efficiency to determine the understanding of the study participants regarding the decision making of an online RL agent.
Thereby, we follow the approach of a ``human-grounded evaluation'' proposed for the evaluation of explainable AI~\cite{Doshi-VelezK17}.

\item \emph{RQ2: What is the perceived usefulness and ease of use of XRL-DINE?} 
To assess the potential acceptance of XRL-DINE, we were interested in the participants' perceived usefulness and perceived ease of use of XRL-DINE.
To do this, we use the TAM (Technology Acceptance Model, see below).
\end{itemize}

While RQ1 focuses on the participants' understanding in regard to the decision making of the online RL agent, RQ2 focuses on the acceptance of XRL-DINE.
Insights on how we designed the user study to measure the participants' task performance and acceptance of XRL-DINE are presented below.

\subsection{Study Design}
\label{sec:s-design}

The overall aim of our user study is to analyze to what degree XRL-DINE helps software engineers to understand the decision making of an online RL agent.
As we extensively motivated in Sect.~\ref{sec:Introduction}, a software engineer cannot understand the decision making of an online RL agent without additional support. 
This was the reason why we introduced XRL-DINE in the first place.
One key design decision that follows from this reasoning is that we opted against a comparative study.
If we compared the participants' performance by splitting into groups with and without XRL-DINE, we would design a rather unfair user study.
Or phrased differently, if we were to compare the performance of two of such groups, we would only measure the degree of unfairness.
The participants in the group without XRL-DINE would have no realistic chance to understand the decision making and thus no realistic chance to solve the tasks correctly without guessing.
We further discuss this design decision in Sect.~\ref{sec:validityRisks}.

One further design decision was to set up the user study in the form of an online questionnaire\footnote{We used the following online survey provider: \url{https://www.limesurvey.org/}}.
This facilitates repeatability (i.e., each participant receives always the same form of questions), simplicity (i.e., no need for filling and scanning paper-based results), and scalability (i.e., easy to share and distribute to additional participants).

Below, we first elaborate on how we designed the study to answer our individual research questions and then provide an overall description of the only questionnaire, followed by insights from pre-testing the study.

\subsubsection{RQ1: Task Performance} 
Following the software engineering literature (e.g., see ~\cite{CeccatoMMNT15,PetersenRW08}), we measure human task performance in terms of effectiveness and efficiency as follows:

\begin{equation}
  \textrm{effectiveness} = \frac{\textrm{Number of correctly performed tasks}}{\textrm{Number of all tasks}} = \frac{\sum c_i}{n}
  \end{equation}

\begin{equation}
\textrm{efficiency} = \frac{\textrm{Number of correctly performed tasks}}{\textrm{Time for performing tasks [Minutes]}}  = \frac{\sum c_i}{\sum t_i}
  \end{equation}

\vspace{1em}

Here, $c_i$ is equal to 1 if the $i$-th task ($i=1,\ldots,n$) was performed correctly, and 0 otherwise, while $t_i$ is the time spent (in minutes) for performing the $i$-th task.

We use the facilities of our online survey provider to measure the time spent for performing each task.
This time includes 1) reading the task, 2) finding the solution, and 3) submitting the solution.
Before facing each task, a disclaimer text makes the participants aware of the upcoming task and time measurement. 
After finishing the task, the participants are informed that time measurement has stopped.

The tasks of the user study were to be performed in the context of the scenario (SWIM exemplar) presented in Sect.~\ref{sec:application}.
To this end, participants were provided with access to the interactive XRL-DINE dashboard\footnote{available at \url{https://fmfeit.github.io/xrl_dine_example_2/}}.
In particular, they were given an excerpt of the scenario covering the 21 time steps after the initial convergence of the RL process.
To reduce the complexity in interacting with XRL-DINE and to facilitate comparability of the performance of different participants, we decided to fix the thresholds $\rho$ (threshold  for ``Uncertain Action'' DINEs) and $\phi$ (threshold for ``Reward Channel Extremum'' DINEs).
We left these values at their default as defined in the prototypical implementation, i.e., $\rho = 0.3$ and $\phi = 0.1$.
We further discuss this design decision in Sect.~\ref{sec:validityRisks}.

\begin{figure*}[ht]
    \centering
	\includegraphics[width=.8\textwidth]{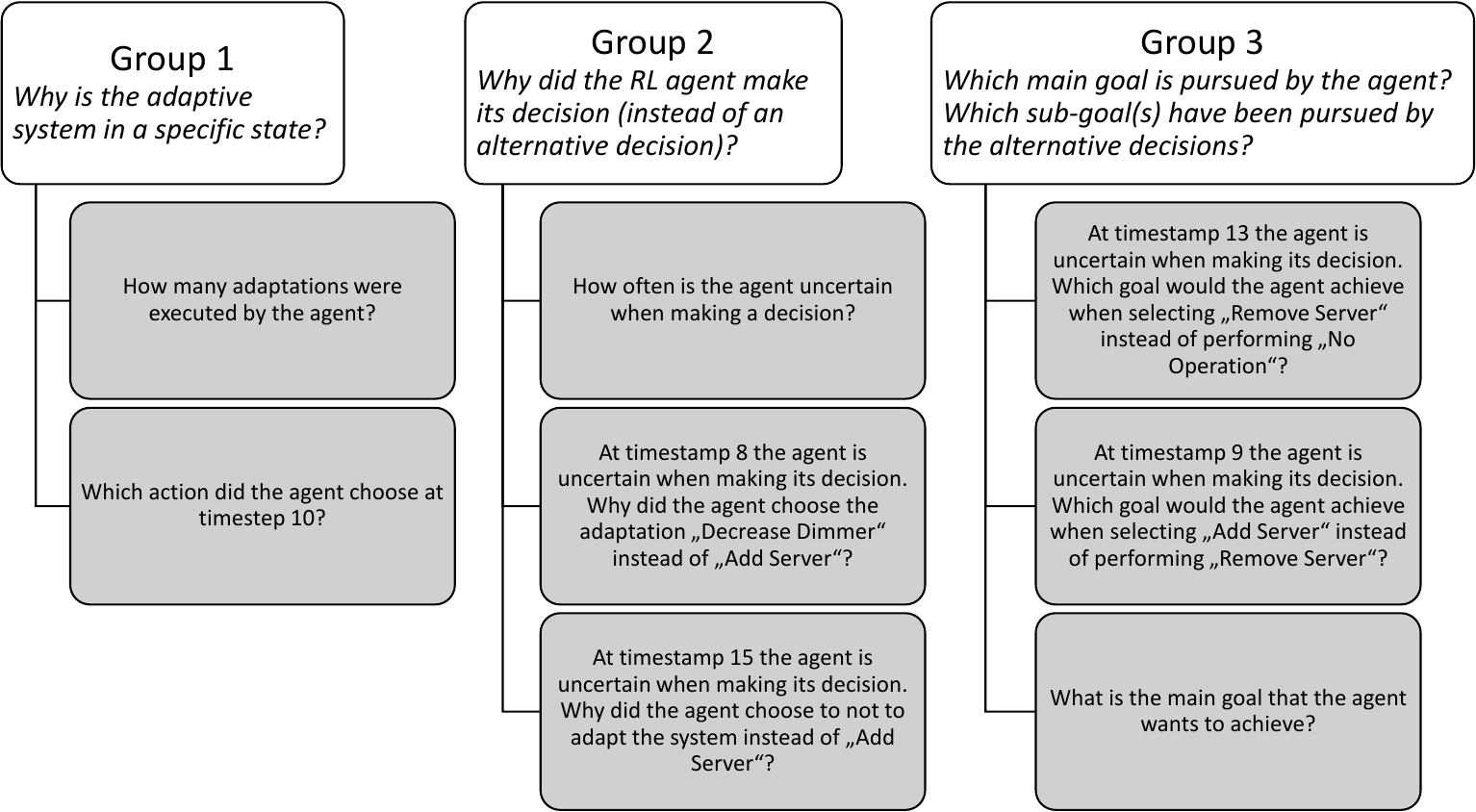}         
   	\caption{Tasks for measuring effectiveness and efficiency}
	\Description[<short description>]{<long description>}
    \label{fig:tasks}
\end{figure*}

Overall, we asked the participants to perform eight tasks clustered into three groups, which are given in Fig.~\ref{fig:tasks} and elaborated below.
The tasks were developed to cover the different aspects that may be answered when using XRL-DINE and incorporate feedback from pre-tests (see Sect.~\ref{sec:pretest}).
The participants had to solve the tasks and provide their answer by answering single-choice questions.
The possible answers are ordered randomly for each participant in order to avoid response bias and order effects.

\runin{Group 1}
Tasks in this group are about determining why the adaptive system is in a specific state. 
In particular, participants had to differentiate between actions of the online RL agent (i.e., every decision made by the online RL agent, which also includes ``No Adaptation''; e.g., see Fig.~\ref{fig:scenario1}) and adaptations (i.e., RL decisions that lead to a change in the system).
	
\runin{Group 2} 
Tasks in this group are about determining why the online RL agent made a particular decision at a given time step and how certain the agent was about this decision, instead of making a possibly alternative decision.
In particular,  participants had to grasp that the online RL agent may sometimes be rather uncertain, and thereby identify possible alternative decisions.
	
\runin{Group 3}
Tasks in this group are about determining which overall goal, as well as which sub-goal is pursued by the online RL agent, and what different sub-goals may have been pursued by the alternative decisions. 

\subsubsection{RQ2: Usefulness and Ease of Use} 
To measure the potential acceptance of XRL-DINE, we used the Technology Acceptance Model (TAM)~\cite{davis1989perceived}.
TAM has been widely studied and emerged as a de-facto standard for analyzing technology acceptance~\cite{marangunic2015technology}.
Advantages of TAM are its low complexity, empirically validated measurement scales, as well as its robustness~\cite{king2006meta, morris1997user}.
Research has also shown that the TAM can be used to evaluate software prototypes \cite{morris1997user, laitenberger1998evaluating, davis2004toward}.

We use TAM to measure the participants' perceived usefulness and perceived ease of use of XRL-DINE.
Perceived usefulness gives insights into how useful participants' perceive XRL-DINE in supporting them to perform tasks related to the decision making of an online RL agent.
Perceived ease of use provides in indicator about how usable, i.e., easy to use XRL-DINE was perceived.

\begin{figure*}[ht]
    \centering
	\includegraphics[width=.65\textwidth]{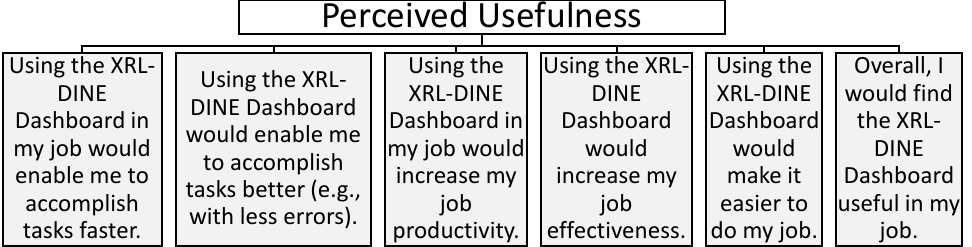}     
\\
	\vspace{1em}
	\includegraphics[width=.65\textwidth]{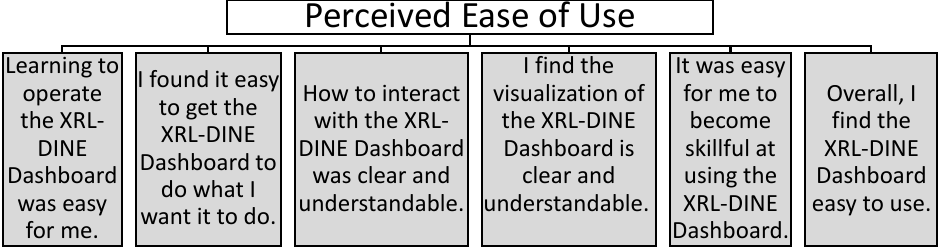}     
	\caption{TAM questions}
	\Description[<short description>]{<long description>}
    \label{fig:tam}
\end{figure*}

Fig.~\ref{fig:tam} shows the TAM questions that we asked the participants.
These questions are based on the typical TAM questions and were slightly modified to fit the scope of the user study, as well as to incorporate feedback from pre-tests (see Sect.~\ref{sec:pretest}).
Following TAM, the answers to each question could be given on a 5-point-scale: 
\textit{extremely unlikely}, \textit{quite unlikely}, \textit{neither}, \textit{quite likely}, and \textit{extremely likely}.

The TAM questions were asked after the participants performed the tasks defined for RQ1.
This means participants were familiar with using XRL-DINE.

\subsubsection{Overall Structure of the Online Questionnaire}
We augmented the parts of the questionnaire related to answering the research questions with additional information, to provide the participants with background information, as well as to determine the participants experience and demographics.

Overall, the questionnaire covered the following parts:

\runin{Part 1} A welcoming-page introduces the study participants to the context of the study.
The participants are asked to adopt the role of a software engineer who seeks to understand the decisions made by an online RL agent.

After this introduction to the study, the participants were asked to rate their experience in machine learning and reinforcement learning, as well as with self-adaptive systems, cloud computing and the XRL-DINE approach.
We used a 4-point scale including \textit{no experience}, \textit{some experience}, \textit{medium experience}, and \textit{high experience}.

\runin{Part 2} 
Part two introduces the participants to the SWIM exemplar, the concrete scenario, as well as XRL-DINE.
The explanatory text presented in this part is available to the participants throughout the whole survey for reference.
The participants are given the following pieces of information: 
\begin{itemize}
	\item An explanation of the SWIM exemplar;
	\item A list of the sub-goals, among which the online RL agent should seek a trade-off;
	\item A list of actions available to the online RL agent together with examples of typical effects of these actions.
\end{itemize}

Note that we did not provide the participants with the reward functions (as introduced in Sect.~\ref{sec:swim}), so the participants had to perform the tasks -- particularly the ones in Group 3 -- solely using XRL-DINE.
When providing the reward function, the participants would be able to identify the weight of each decomposed reward channel.
However, we want to measure among others whether the explanations of XRL-DINE are sufficient to identify the main  goal pursued by the online RL agent/the sub-goal pursued when selecting an alternative decision (see Group 3 of the tasks).
Thus, we did not provide the participants with the reward functions.

\begin{figure*}[ht]
	\centering
	\includegraphics[width=.7\textwidth]{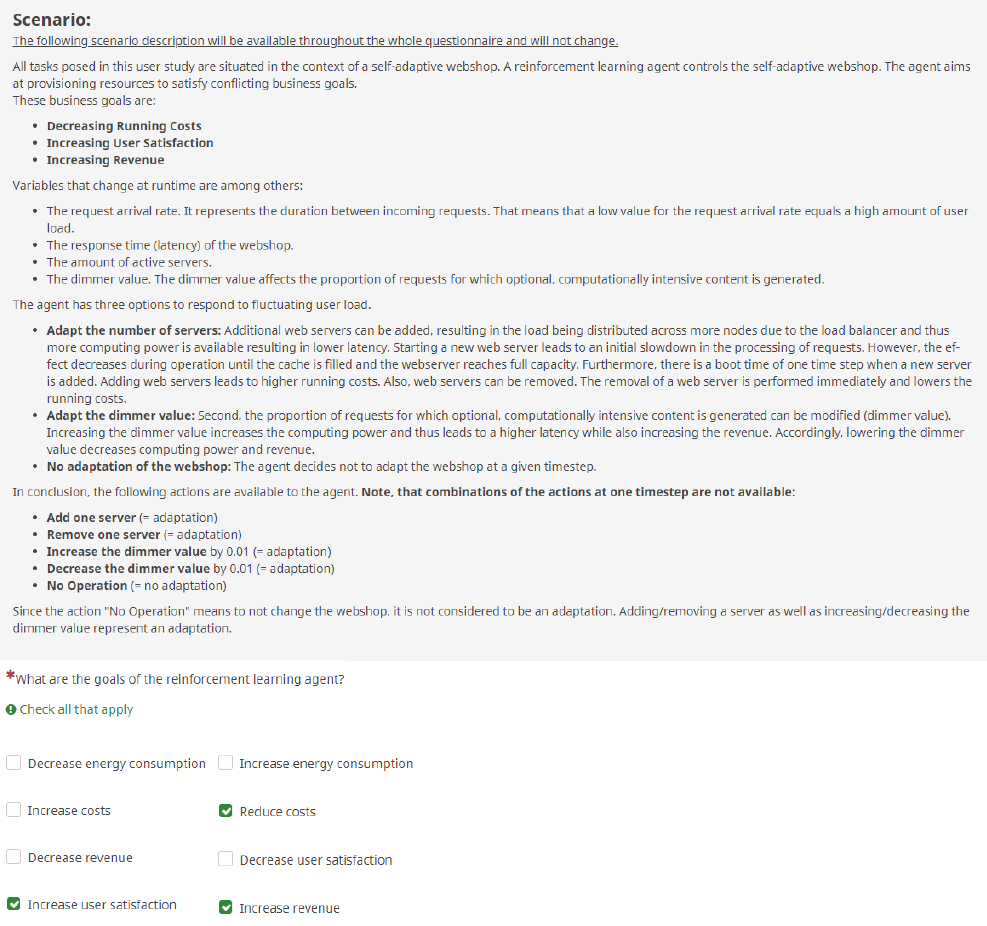}     
	\caption{Screenshots of the online questionnaire showing the description texts and a selected associated control question}
	\Description[<short description>]{<long description>}
	\label{fig:descriptionTexts}
\end{figure*}

After this introduction to the exemplar, participants are asked control questions to check in how far participants read and ``understood'' the information provided.
They ask, e.g., about the goals of the online RL agent, available actions, and which part of the dashboard provides which kind of information.
Control questions were posed as multiple-choice questions with different textual answers.
An excerpt of the online questionnaire showing the description texts and a selected associated control question is shown in Fig. \ref{fig:descriptionTexts}.

\runin{Part 3} 
This part of the study focuses on measuring task performance as elaborated for RQ1 above.
%An example from the online questionnaire is shown in Fig \ref{fig:task7}

After completing each task, participants are asked to indicate which part(s) of the XRL-DINE dashboard they used to solve the task.
A multiple-choice list including each part of the dashboard is given.

%\begin{figure*}[ht]
%	\centering
%	\includegraphics[width=1\textwidth]{fig/Task 7 Screenshot.png}     
%	\caption{Screenshot of the online questionnaire including Task 7}
%	\Description[<short description>]{<long description>}
%	\label{fig:task7}
%\end{figure*}

\runin{Part 4} 
This part addresses RQ2 by asking TAM questions as explained above.
%Fig \ref{fig:tam_screenshot} shows an extract from the online questionnaire including the TAM questions regarding the participants' perceived usefulness.

%\begin{figure*}[ht]
%	\centering
%	\includegraphics[width=1\textwidth]{fig/TAM_Screenshot.png}     
%	\caption{Screenshot of the online questionnaire including the TAM questions regarding perceived usefulness}
%	\Description[<short description>]{<long description>}
%	\label{fig:tam_screenshot}
%\end{figure*}

\runin{Part 5} 
This part served to assess in how far participants faced problems during the study.
Participants are asked open-ended questions regarding the descriptions texts (scenario and dashboard description) and/or the visualization of the XRL-DINE dashboard. 
Participants were also encouraged to provide ideas on how to improve the XRL-DINE dashboard.

\runin{Part 6} 
This final part collects further demographic data from participants, including gender, age group, highest degree, and job.
%We also asked for the type of end device they used to participate in the study.

\subsubsection{Pre-testing}
\label{sec:pretest}
Before finalizing the questionnaire, we conducted a pre-testing with five researchers.
Pre-testing included (1) proof-reading of the text, i.e., scenario and dashboard description, intro, and questions, (2) answering the questions and working on the tasks while thinking out loud, (3) observing how the XRL-DINE dashboard was used, and (4) a concluding open discussion.

Pre-testing led to improvements in the description texts to improve understanding of the scenario and dashboard parts as well as to improve readability. 
For example, highlights were added to texts providing reader guidance.
For what concerns RQ1, the wording of the task was refined and made less ambiguous.
For what concerns RQ2, questions of the TAM questionnaire were slightly revised to better fit the situation at hand and reduce ambiguity. 
%For example, ``Using the XRL-DINE dashboard would improve my job performance'' has been changed to ``Using the XRL-DINE Dashboard would enable me to accomplish tasks better (e.g., with less errors)'' to define the meaning of performance. 

Pre-testing also allowed us to estimate the difficulty of each task.
Moreover, we were able to determine an approximate completion time for the survey.
%Last but not least, by discussing the content of the survey, we were able to strengthen the alignment of the survey and the survey goals.

\subsection{Study Execution}
\label{sec:s-execution}
We executed the user study in a time frame of 30 days (26.01.2023 -- 24.02.2023).
All participants were directly contacted by the second author (via e-mail or in person).
As participants, we approached software engineers from academia and industry according to the criteria defined for a ``human-grounded evaluation'' in~\cite{Doshi-VelezK17}. 
In particular this means, as we perform simplified tasks, we could approach participants that were knowledgeable in software engineering, but not necessarily had to be experts in the fields covered by our study. 

We collected 54 fully completed questionnaires and 23 partially completed questionnaires.
For further analysis, only the completed questionnaires were taken into account, regardless of whether the participants answered the control questions correctly.

Tab.~\ref{Tab:demographics} provides an overview of the demographics of the study participants.
%The participant pool is dominated by male participants (81\%).
The dominant age group is 20-29 (57\%) followed by the age group 30-39 (24\%).
Most participants hold a Bachelor's or Master's degree (76\%).
While we asked both, people from academia and industry to participate in the study, most participants can be assigned to the group academia including researchers and students (82\%).
%Most participants followed our appeal and used a computer or a laptop (92\%). Only 8\% chose to use a smartphone or tablet.

\begin{table}[h]
	\caption{Demographics of the participant pool}
	\label{Tab:demographics}
	\begin{tabular}{cccccc}
		\toprule
%		\multirow{2}{*}{Gender}     &  Male     & Female & Diverse    & Prefer not to say &       \\ 
%		& 81\%     & 9\%    & 0\%        & 9\%               &       \\ \midrule
		\multirow{2}{*}{Age}        & 20-29    & 30-39  & 40-49      & 50-59             & 60-69 \\ 
		& 57\%     & 24\%   & 7\%        & 7\%               & 4\%   \\ \midrule
		\multirow{2}{*}{Degree} & High school graduate & Bachelor's degree & Master's degree & PhD      & Prefer not to say \\ 
		& 11\%     & 35\%   & 41\%       & 11\%              & 2\%   \\ \midrule
		\multirow{2}{*}{Job}    & Bachelor Student     & Master Student    & Researcher      & Industry & Prefer not to say \\ 
		& 17\%     & 28\%   & 37\%       & 11\%              & 7\%   \\ %\midrule	\multirow{2}{*}{End Device} & Computer & Laptop & Smartphone & Tablet            &       \\ 
		%& 28\%     & 65\%   & 6\%        & 2\%               &       \\ 
		\bottomrule
	\end{tabular}
\end{table}

\subsection{Study Results}
\label{sec:s-results}

Below we present the results of our user study for each of the research questions. 

\subsubsection{Results for Task Performance (RQ1)} 
\begin{figure*}[ht]
	\centering
	\includegraphics[width=.8\textwidth]{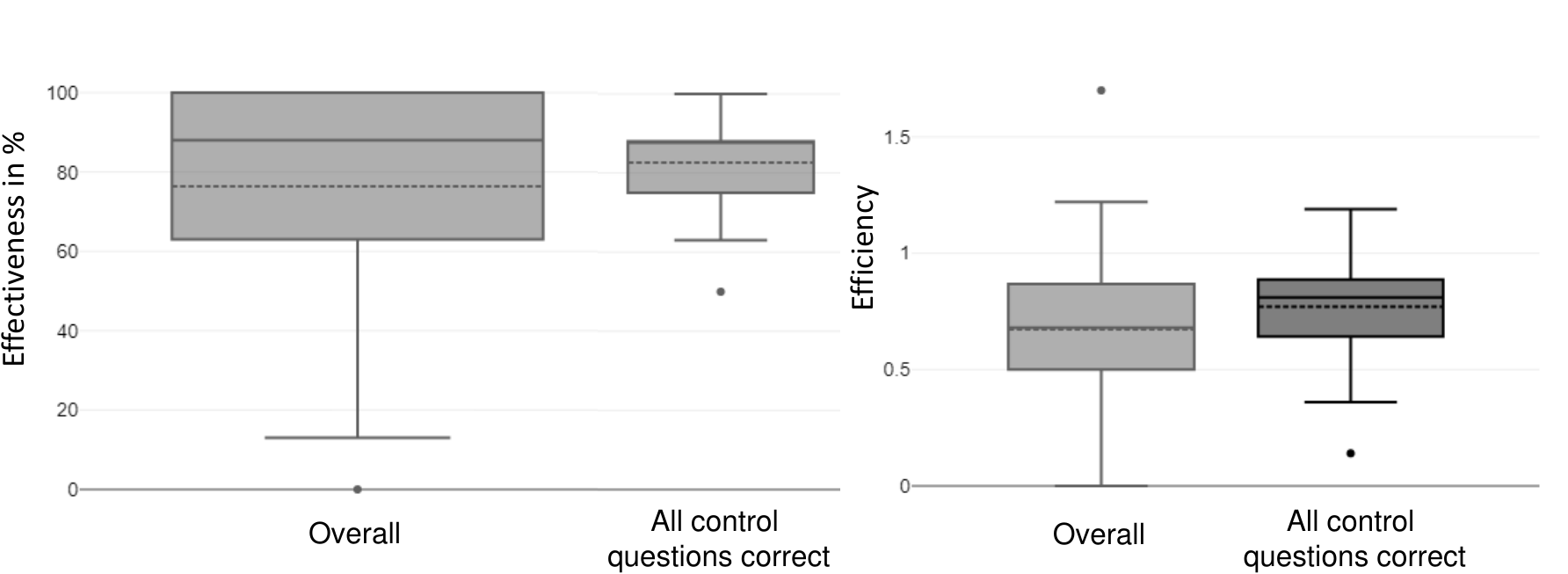}	
	\caption{Effectiveness and Efficiency (dotted line = mean; bold line = median)}
	\label{fig:overallTaskEffectivenessAndEfficiency}
\end{figure*}

Tab. \ref{Tab:solutionPerTask} shows the participants' mean performance per task.
Fig. \ref{fig:overallTaskEffectivenessAndEfficiency} shows effectiveness and efficiency as boxplots to depict the spread of performance across participants.
Two sets of boxplots are shown: (1) for the overall set of participants ($n = 54$) and, (2) only for the participants that answered the control questions correctly ($n = 22, 41\%$ of all 54 participants).

\begin{table}[h]
	\caption{Participants' mean performance per task}
	\begin{tabular}{@{}llllllllll@{}}
		\toprule
		& Task 1 & Task 2 & Task 3 & Task 4 & Task 5 & Task 6 & Task 7 & Task 8 & Mean\\ \midrule
		Number of correct answers & 35 & 49 & 51 & 33 & 34 & 43 & 37 & 45 & 40.875\\
		Time to answer [Min:Sec] & 02:00   & 00:56    & 00:47    & 01:59   & 01:30   & 01:05   & 00:46   & 01:16 & 1:17  \\ 
\midrule
		Effectiveness & 65\%   & 91\%   & 94\%   & 61\%   & 63\%   & 80\%   & 69\%   & 83\%  & 76\% \\
		Efficiency & 0.32   & 0.97 & 1.21   & 0.31    & 0.42    & 0.73    & 0.89    & 0.66  & 0.67  \\
		\bottomrule
	\end{tabular}
	\label{Tab:solutionPerTask}
\end{table}

%% ZUSTIMMUNG ;) Weglassen
%\todo{@Andreas: Ich habe mir die beiden Paper nochmal angesehen und denke nicht, dass es sinnvoll ist diese für den Kontext zu verwenden. Ich suche gleich nochmal nach Papern die Efficiency bei Tasks genutzt haben.}
%\todo{contextualize efficiency; e.g., Cf. Mean efficiency of tool-supported debugging: 0.03 corrected faults / minute \cite{CeccatoMMNT15};  mean efficiency of requirements and design inspection 0.067 defects per minute \cite{PetersenRW08}}

\runin{Effectiveness}
On average (across all eight tasks), participants achieved an effectiveness of 76\%, which means that they correctly performed more than $3/4$ of all tasks.
Of all 54 participants, 33\% were able to solve all eight tasks correctly, while only 2\% failed to correctly answer any of the task.
This means that there were few outliers who could not solve any of the tasks while the majority could solve most tasks with the help of  XRL-DINE.
When looking at the different task groups one can see that the participants' achieved the highest effectiveness for Group 1 (80\%), followed by Group 3 (77\%) and Group 2 (72\%).
It can be concluded that the participants achieved similar results for all three task groups, although they identified most effectively why the adaptive system is in a specific state.
When comparing the effectiveness of the participants with giving a random answer (mean effectiveness of giving a random answer equals 21\%), we calculated an effect size of $r = 0.82$ using the non-parametric Mann-Whitney U-Test at a significance level of $p = <.001$.

When considering participants who gave the correct answers to all control questions only, we can measure a 6\% higher mean effectiveness of 82\%.
The Mann-Whitney U-Test indicates an effect size of $r = 0.92$ with $ p  < .001$ when comparing to participants giving a random answer.
Filtering for correct answers to the control questions has especially removed the outliers and reduced the standard deviation. 

\runin{Efficiency}
The average efficiency was 0.67, which means that it took the participants a little over $2/3$ of a minute for each correct task.
The highest efficiency was achieved by a participant who completed all tasks correctly in under five minutes (efficiency = 1.7).
The lowest efficiency equals 0, since some participants could not solve any of the tasks correctly.
Looking at the three groups of task, the participants achieved similar efficiency for Group 1 (0.77) and Group 2 (0.76).
The participants' efficiency solving tasks from Group 2 is noticeably worse (0.57).
This is because the participants took considerably longer to find a solution to tasks 4 and 5, which reduces the participants' efficiency solving tasks from Group 2.
Thus, identifying why the agent made its decision instead of an alternative decision may be not as straight forward (in terms of efficiency) as identifying the main-/sub-goal(s) of the online RL agent or identifying why the adaptive system is in a specific state.

When filtering the participants based on their answers to the control questions, one can identify fewer outliers and a 15\% higher mean efficiency of 0.77.

% natürlich korreliert das, die Formeln haben ja beide denselben zaehler ;-) Weglassen!
%A Spearman correlation was performed to test whether there was an association between Effectiveness and Efficiency. 
%The result of the Spearman correlation showed that there was a significant association between effectiveness and efficiency, $r = 0,58, p = <.001$.
%Hence, effectiveness and efficiency together form a suitable picture for the participants' task performance.

\runin{Analysis by degree and job level}
As we collected the participants' degree and job levels (see Tab.~\ref{Tab:demographics}), we also analyzed whether there was a significant difference between the task performance of different groups of participants.
As an example, one would expect a participant with a higher degree to outperform less experienced participants with a lower degree.

We again used the Mann-Whitney U-Test to measure the effect size between the various groups.
However, we could not measure any statistically significant difference between the groups, so our data does not allow us to conclude whether participants' degree and job levels affect task performance.

\subsubsection{Results for Usefulness and Ease of Use (RQ2)} 
\begin{figure*}[ht]
	\centering
	\includegraphics[width=0.5\textwidth]{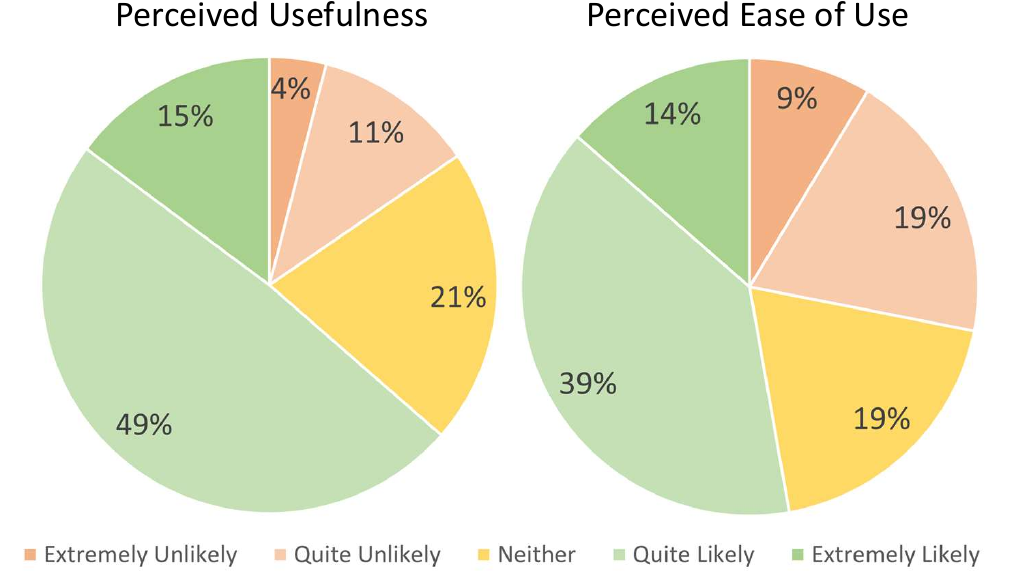}	
	\caption{Overall Results for Perceived Usefulness and Perceived Ease of Use}
	\label{fig:tamTorte}
\end{figure*}

Fig.~\ref{fig:tamTorte} shows the accumulated results for the TAM questions as pie charts.

As can be seen, the participants rated their perception regarding the usefulness and ease of use of the XRL-DINE dashboard positively, i.e., 64\% of all statements regarding the participants' perceived usefulness was given a rating of \textit{Quite Likely} or \textit{Extremely Likely}. 
Only 15\% perceive XRL-DINE as not being useful.
Perceived ease of use came off slightly worse than perceived usefulness, but overall the positive ratings outweigh the negative ones here as well (53\% positive vs. 28\% negative).

\begin{figure*}[ht]
	\centering
	\includegraphics[width=.9\textwidth]{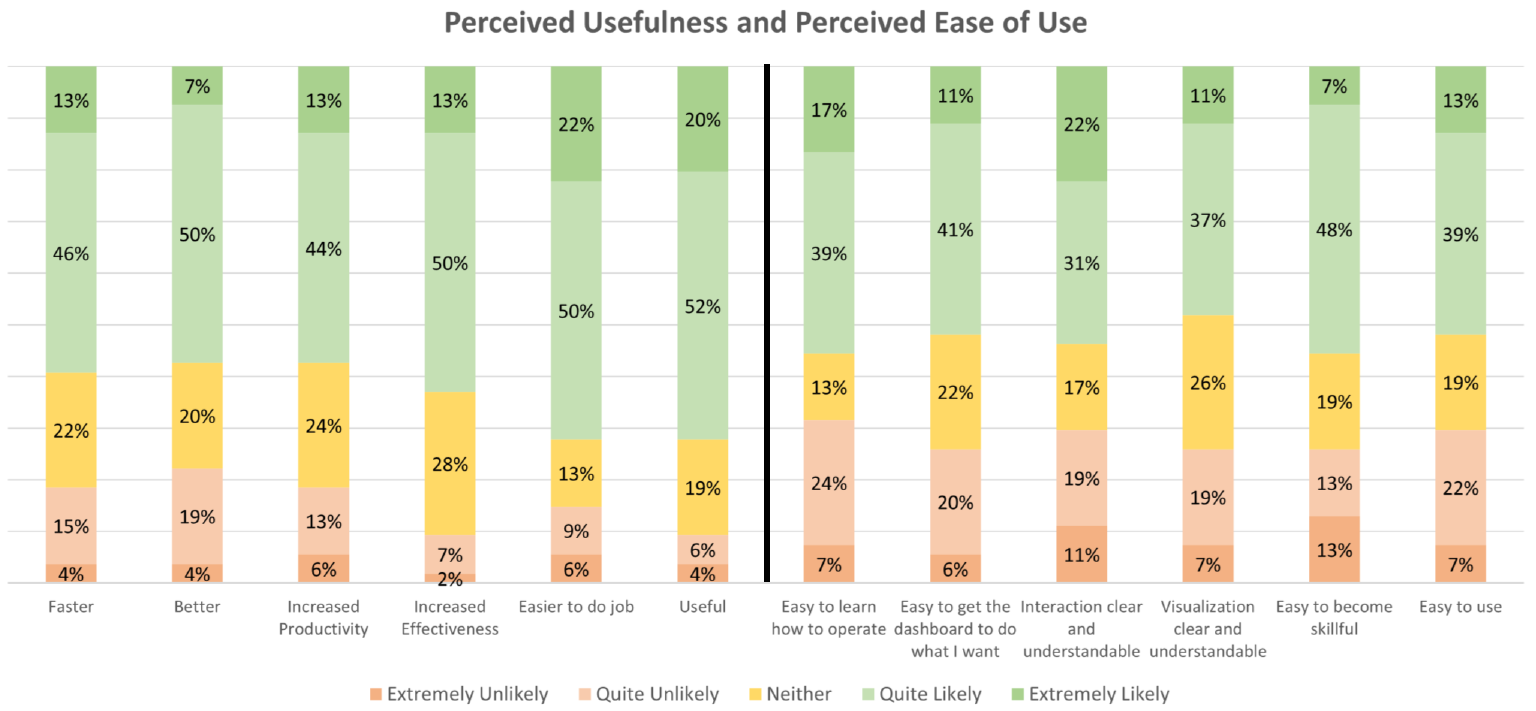}	
	\caption{Results for each Question of the TAM Questionnaire}
	\label{fig:tamResults}
\end{figure*}

Fig.~\ref{fig:tamResults} shows a breakdown per individual TAM question as bar charts.
Each bar provides the relative number of answers to each of the TAM questions.
As can be seen, all questions were rated with at least 57\% of positive votes.
The best results were achieved for the statement ``Using the XRL-DINE Dashboard would make it easier to do my job.'' (72\% positive ratings vs. 10\% negative ratings).

Yet, more than a quarter of all participants stated challenges when using XRL-DINE, i.e., issues with understanding the visualization and problems getting the dashboard to do what they wanted.
This result also goes hand in hand with the feedback from the free-text fields, where the visualization of the dashboard is criticized by several participants.
Also, almost a third of all participants had problems to learn how to operate the XRL-DINE dashboard (31\% negative ratings for the respective statement).
Hence, although the perceived usability of XRL-DINE has been positive on average, there is potential for improvement, which we thus discuss in Sec.~\ref{sec:discussion}.

\section{Discussion} 
\label{sec:discussion}

Here we analyze validity risks of the empirical results presented in Sect.~\ref{sec:application} and~\ref{sec:user-study}, present the reproducibility of our findings, and discuss potential enhancements of XRL-DINE.

\subsection{Validity Risks}
\label{sec:validityRisks} 
Our user study inherits the typical validity risks of user studies on task performance (e.g., see~\cite{DaunBOS21}).
We discuss them along the different types of risks below.

\runin{Internal Validity}
There is a risk that participants may not seriously participate or may try to over-perform. 
To address this risk, we designed the study as an online questionnaires to be conducted within 30--45 minutes, thereby reducing the risk that participants may lose interest.
Also, we did not offer the option of pausing and returning to the questionnaire to avoid participants spend arbitrary long times for answering the questions. 

Concerning the answer choices for RQ2, we used verbal scale labels, as numeric labels might convey the wrong meaning. 
We also started with the negative options first, as studies have shown that people tend to select the first option that fits within their range of opinion. 
By doing so, our study thus leads to more “conservative” results (cf.~\cite{MetzgerM11}).

\runin{Construct Validity}
We designed the survey in such a way that it provides each participant with a detailed introduction and thus equal grounding for answering the questions.
Thereby, we avoid the risk that participants would just guess the answers, which is confirmed by measuring the effect size when comparing the answers given with random answering.

As we justified above, we decided against a comparative study where we would compare participants that have access to XRL-DINE to participants that would not have access, as this would be an unfair comparison.
Our results indicate that XRL-DINE indeed helps software engineers understand the decision making of an online RL agent.
So, follow-up work may provide additional, more detailed analysis of what elements of XRL-DINE indeed have which effect on this positive results.
As an example, one could compare the use of different DINEs, or even compare XRL-DINE against its underlying two baseline techniques.

\runin{External Validity} We used an actual adaptive system exemplar together with 30 GB of real-world workload traces when validating XRL-DINE.
We tuned the hyperparameters of the RL algorithm experimentally, using educated guessing (e.g., comparable to~\cite{mnih_human-level_2015}). 
We purposefully did not perform extensive, exhaustive hyperparameter tuning, e.g., using grid search, because our aim was not to improve or compare the performance of existing RL approaches, but to validate how XRL-DINE may be used to generate explanations for RL decisions.
Still results are only for a single system, which thus limits generalizability.

We used participants with various degrees and job types, providing a cross-section of some typical software engineering personnel.
We could measure a positive effect of using XRL-DINE for all participants, however it appears that we could not measure a statistically significant effect across the different degrees and job types.
If we were to generalize our findings for specific types of software engineers, an enlargement of the user study thus would be needed.

\subsection{Reproducibility}
\label{subsec:repro} 
To facilitate reproducibility of our research, we provide relevant background and supplementary material in an online repository at: \url{https://git.uni-due.de/rl4sas/xrl-dine}.
This material includes the prototypical implementation (code) for XRL-DINE, the setup of the SWIM exemplar (including the interface wrapper as well as the generated workload patterns), the running XRL-DINE dashboards, as well as the data of the user study.

\subsection{Potential Enhancements of XRL-DINE} 
\runin{Addressing multidisciplinary requirements}
Addressing relevant requirements stemming from multidisciplinary research could address some of the problems with respect to usability perceived by the study participants.

In particular, \tcite{miller_explanation_2019} presents the results of multidisciplinary research, identifying four relevant requirements that explanations should exhibit:
\begin{itemize}
\item Explanations should be \emph{contrastive}, i.e., people seek counterfactual cases, meaning that they do not ask why event $P$ happened, but rather why event $P$ happened instead of some event $Q$.
XRL-DINE provides contrastive explanations in the form of the Reward Channel Dominance DINEs as well as the textual counterfactuals generated for the Uncertain Interaction DINEs.

\item Explanations should be \emph{selected}.
XRL-DINE provides selective explanations in the form of Uncertain Action DINEs, as well as by being able to tune the rate of DINEs shown.

\item Explanations should be \emph{causal}, i.e., providing true causes for an observed event is more effective than giving statistical relationships.
XRL-DINE is not yet causal.
Work by~\tcite{Dusparic_2022} indicates a potential direction along which XRL-DINE may be extended to become causal.
They propose checking whether the RL agent relies on spurious correlations (and thus not on causality).

\item Explanations should be \emph{conversational}, i.e., presented as part of a conversation (as explained in Sect.~\ref{sec:Foundations}, explanation entails a process involving the explainer and the explainee).
While the XRL-DINE dashboard facilitates interaction between the explainee (XRL-DINE user) and the explainer (XRL-DINE), this interaction does not take the form of a conversation.
Here, XRL-DINE may be enhanced by leveraging natural language understanding and generative AI to provide conversational explanations; e.g., via large language models and the concept of chatbots~\cite{GaoLXA21}.
\end{itemize}

\runin{Explainability in the presence of delayed rewards}  
Environment dynamics may delay the effects of adaptations. 
As an example, in the SWIM exemplar (introduced in Sect.~\ref{sec:application}) executing the \emph{Add Server} action requires booting a new server, which takes time.
As a result, the effect on latency and thus on rewards for executing this adaptation are delayed.
In contrast, lowering the dimmer value has an immediate effect on latency and thus on rewards. 
Such timing-related differences may make interpreting DINEs more difficult.
To address these timing-related differences, XRL-DINE may for instance be extended by the RUDDER approach for decomposition of delayed rewards~\cite{Arjona-MedinaGW19}.

\runin{Explanations for decentralized adaptive systems} 
In a decentralized adaptive systems, the adaptation logic is decentralized across multiple systems~\cite{QuinWG21,DAngeloGGGNPT19}.
As XRL-DINE is built to explain the decisions of a single RL agent, XRL-DINE does not consider the decisions of other RL agents during explanations.
Similarly to the aforementioned situations in which XRL-DINE may generate difficult to understand DINEs, the same may happen if XRL-DINE is directly applied to decentralized adaptive systems.
Extending XRL-DINE depends on the fundamental, underlying approach of decentralization.
On the one hand, there are different shapes that such a decentralization may take, i.e., different shapes of how the MAPE-K elements are decentralized across systems~\cite{QuinWG21}.
On the other hand, there are different ways of how to decentralize Online RL, including Multi-Agent RL~\cite{MoustafaZ14}, hierarchical RL~\cite{CaporuscioDGM16}, and meta RL~\cite{ZhangLZTHJ21}.

\runin{Explanations for policy-based RL to capture concept drifts}
XRL-DINE works for value-based Deep RL (see Sect.~\ref{sec:Foundations}), because XRL-DINE needs access to the learned action-value function $Q(S,A)$.
%While value-based RL can cope with multi-dimensional, continuous state and action spaces, as well as generalize over unseen neighboring states, value-based RL faces the exploration-exploitation dilemma~\cite{sutton_reinforcement_2018}. 
%Actions should be selected that have shown to be effective (aka. exploitation). 
%However, to discover such actions in the first place, actions that were not selected before should be selected (aka. exploration). 
%One typical solution to the exploration-exploitation dilemma is the $\epsilon$-greedy mechanism. 
%During learning, a random action is chosen with probability $\epsilon$ (exploration), while the action with the highest expected cumulative reward as determined by $Q(S,A)$ is chosen with probability $1-\epsilon$ (exploitation). 
%The challenge for a software engineer  is to fine-tune the balance between exploitation and exploration to ensure
%convergence of the learning process~\cite{CAISE2020}.
%As an example, one may implement a mechanism that decreases $\epsilon$ over time, thereby reducing the amount of exploration  to facilitate convergence. 
As explained in Sect.~\ref{sec:Approach}, value-based RL typically relies on $\epsilon$-greedy strategies to address the exploration-exploitation dilemma. 
This poses the additional challenge of when and how to increase $\epsilon$  again
in order to capture concept drifts in the system environment. 
Concept drifts occur due to an evolution of the system environment that lead to a change of the effects of adaptations.
As an example, if the physical machines that provide the virtual servers in the SWIM exemplar would be replaced by less powerful, but more energy efficient machines, this would impact on the effect of the \emph{Add Server} action  in terms of latency.
%As such changes of physical machines are not monitored by the SEAMS exemplar, the RL agent is not able to observe this change by considering the state space $S$.
Coping with such concept drift in value-based Deep RL would require to observe such concept drift and to increase the exploration rate to learn the changed effect of adaptations.

Policy-based Deep RL~\cite{NachumNXS17,SuttonMSM99} can naturally cope with such concept drifts~\cite{CAISE2020,BPM2020}.
The fundamental idea of policy-based RL is to directly use and optimize a parametrized stochastic action selection policy $\pi_\theta(S,A)$ in the form of a deep artificial neural network.
The action selection policy $\pi_\theta(S,A)$ gives the probability of taking adaptation action $A$ in state $S$, i.e., $\pi_\theta(S,A) = \mathrm{Pr}(A|S)$.
Using stochastic action selection, policy-based RL captures concept drifts of the system environment without the need for software engineers to intervene~\cite{CAISE2020,BPM2020}.

\section{Related Work}
\label{sec:RelatedWork}

While generic explainable RL approaches are discussed in recent overview papers (such as~\cite{heuillet_explainability_2020, puiutta_explainable_2020}), these do not specifically address adaptive systems. 
We thus focus the following discussion on solutions that specifically provide explanations for adaptive systems, which can be clustered into the following main groups:

\runin{Temporal graph models} 
This group of work uses temporal graph models as central artifact to derive explanations~\cite{garcia-dominguez_towards_2019, UllauriGBZZBOY22}. 
As proposed by~\tcite{garcia-dominguez_towards_2019}, such a model may be used for so called forensic self-explanation.
This means that the model may be queried via a dedicated query language. 
In addition, such a model may be used for so called live self-explanation.
Here, one can submit queries to the running system and be presented with a live visualization. 
The underlying temporal model is kept up to date at runtime (i.e., employed as a model@runtime).
The approach is comparable to the Interestingness Elements described in Sect.~\ref{sec:Foundations} since explanations are generated based on execution traces. 
However, in contrast to XRL-DINE, interesting interactions must be extracted by manually writing queries using the provided query language. 
As follow-up work, suggestions for automating the selection of interesting interaction moments are proposed.
~\tcite{UllauriGBZZBOY22} fully automated automate this by using complex-event-processing.
While the aim of \citeauthor{UllauriGBZZBOY22} is to select interesting interactions to keep the size of the models@runtime manageable, the aim of XRL-DINE is to reduce the cognitive load of developers. 
Compared to earlier work on temporal graph models for explainable adaptive systems, ~\cite{UllauriGBZZBOY22} stands out in providing explanations for RL decisions.
While \citeauthor{UllauriGBZZBOY22} hint at the possibility of using model@runtime queries to realize reward decomposition, it differs from XRL-DINE in that the combination of Interestingness Elements and Reward Decomposition is not considered.

\runin{Goal models} 
This group of work uses goal-based models at runtime~\cite{bencomo_self-explanation_2012, welsh_self-explanation_2014}. 
~\tcite{bencomo_self-explanation_2012} use higher-level system traces as explanations.
Again, this can be considered similar to the idea of Interestingness Elements.
~\tcite{welsh_self-explanation_2014} employ a domain-specific language for providing explanations in terms of the satisficement of softgoals. 
In this regard, explanations are comparable to Reward Decomposition explanations introduced by ~\tcite{juozapaitis_explainable_2019} as described in Sect.~\ref{sec:Approach} in that the explanations refer to competing goal dimensions. 
Other than XRL-DINE, these techniques require making assumptions about the environment dynamics at design time, which can be a source for error due to design time uncertainty~\cite{weyns_perpetual_2013}.
Also, in contrast to XRL-DINE, this group of work does not explicitly consider RL.

\runin{Provenance graphs} 
\tcite{reynolds_automated_2020} employs interaction data collected at runtime to generate explanations in the form of provenance graphs. 
A provenance graph contains information and relationships that contributed to the existence of a piece of data. 
By keeping a history of different versions of the provenance graph, it is possible to determine at runtime \textit{if} and \textit{how} the model has changed (using model versioning) and \textit{who} has changed the model and \textit{why} (using the provenance graph).
Provenance graphs quickly can become too complex to be meaningfully interpreted by humans, thus a dedicated query language was introduced that allows extracting information of interest.
Again, in contrast to XRL-DINE, this group of work does not consider RL.

\runin{Anomaly detection}
\tcite{ZiescheKG21} suggest using machine learning to detect anomalous behavior of an adaptive system which may require an explanation.
Again, this is similar to the idea of Interestingness Elements, which allows determining relevant points for explanation. 
In addition, they reduce the typically huge search space of possible reasons for such anomalous behavior by classifying the behavior into classes with similar reasons.
Thereby, their approach can be considered a first step towards  so called "self-explainable" systems that autonomously explain behavior that differs from anticipated behavior. 
Again, in contrast to XRL-DINE, this group of work does not explicitly consider RL.

\runin{Explainability as a tactic}
\tcite{li_hey_2021} take a fundamentally different view on explainability. 
A formal framework is proposed in which explainability is not provided externally but is considered a concrete tactic of the adaptive system. 
In uncertain or difficult situations, the adaptive system can ask assistance from a human operator in making a decision rather than acting itself. 
Specifically, the overall system is modeled as a turn-based, stochastic multiplayer game in which three players participate. 
These players are (1) the actual self-adaptive system, (2) the environment, and (3) the operator. 
This game is then analyzed using a probabilistic model checker to determine when the involvement of a human operator is necessary.
To prevent the operator from being permanently consulted, there is a cost to using this tactic that must be accounted for by the model checker. 
In this respect, this approach is similar to XRL-DINE, which aims to reduce the cognitive burden of the human. 
However, the motivation in XLR-DINE is the limited cognitive ability of the human, while the motivation for~\citeauthor{li_hey_2021} it is the time delay caused by involving an operator.
Also here, RL is not explicitly considered.

\runin{Explainable online reinforcement learning}
In our previous work~\cite{ACSOS22}, we introduced XRL-DINE providing detailed explanations of RL decisions at relevant points in time.
XRL-DINE enhances and combines two existing explainable RL techniques from~\tcite{juozapaitis_explainable_2019} and~\tcite{sequeira_interestingness_2020}.
The explainable RL techniques were proposed in isolation and not tailored to the needs of explaining adaptive systems.
In ~\cite{ACSOS22}, we thus proposed a combination and adaptation of these generic techniques.
In contrast to our previous work, this paper adds the design, execution and analysis of a user study.
On the one hand, we assessed the efficiency and effectiveness of software developers when using XRL-DINE to perform specific tasks related to the debugging of an adaptive system.
On the other hand, we assesses the perceived usefulness and usability of XRL-DINE.

\section{Conclusion and Perspectives}
\label{sec:ConclusionAndPerspectives}

We introduced XRL-DINE, a technique that helps understanding the decision making of Online Deep RL for adaptive systems.
We described the prototypical implementation of XRL-DINE using a state-of-the-art deep RL algorithm, serving as proof-of-concept.
We used an adaptive systems exemplar to demonstrate the use of XRL-DINE, measure indicators for the cognitive load of using XRL-DINE.
In particular, we performed a user study involving 54 software engineers.
Results show that XRL-DINE helps to correctly perform tasks (with a rate of 76\% correctly executed tasks) and that these tasks can be performed in reasonable amount of time (on average $2/3$ of a minute per correct task).
Additionally, XRL-DINE is perceived useful by the majority of participants and considered usable, with indications for future enhancements.

\begin{acks}
We cordially thank the participants of our user study.
Research leading to these results  received funding from the EU's Horizon 2020 and Horizon Europe R\&I programmes under grant agreements 101070455 (DynaBIC) and 871493 (DataPorts).  
\end{acks}

\bibliographystyle{ACM-Reference-Format}
\bibliography{taas,taas-extra}

\end{document}